\documentclass[11pt]{article}
\usepackage[T1]{fontenc}
\usepackage{graphicx}
\usepackage{booktabs}
\usepackage[margin=20mm]{geometry}

\title{Statistical Inference for Ordinal Predictors in Generalized Linear and Additive Models with Application to Bronchopulmonary Dysplasia} 	

\author{Jan Gertheiss$^{1,*}$, Fabian Scheipl$^2$, Tina Lauer$^{3,4}$, Harald Ehrhardt$^{3,4}$}

\begin{document}
	
\maketitle

\begin{center}
$^1$School of Economics and Social Sciences, Helmut Schmidt University, Hamburg, Germany\medskip\\

$^2$Department of Statistics, Ludwig Maximilians University, Munich, Germany\medskip\\

$^3$Department of General Pediatrics and Neonatology, Justus Liebig University, Giessen, Germany\medskip\\

$^4$German Center for Lung Research (DZL), Universities of Giessen and Marburg Lung Center (UGMLC), Giessen, Germany\medskip\\

$^*$To whom correspondence should be addressed: \texttt{jan.gertheiss@hsu-hh.de}\\

\vspace{10mm}
{\large Abstract}
\end{center}

Discrete but ordered covariates are quite common in applied statistics, and some regularized fitting procedures have been proposed for proper handling of ordinal predictors in statistical modeling. In this study, we show how quadratic penalties on adjacent dummy coefficients of ordinal predictors proposed in the literature can be incorporated in the framework of generalized additive models, making tools for statistical inference developed there available for ordinal predictors as well. Motivated by an application from neonatal medicine, we discuss whether results obtained when constructing confidence intervals and testing significance of smooth terms in generalized additive models are useful with ordinal predictors/penalties as well.\\


\section{Introduction}\label{intro}

\begin{figure}[tb]\centering
	\includegraphics[width=57mm]{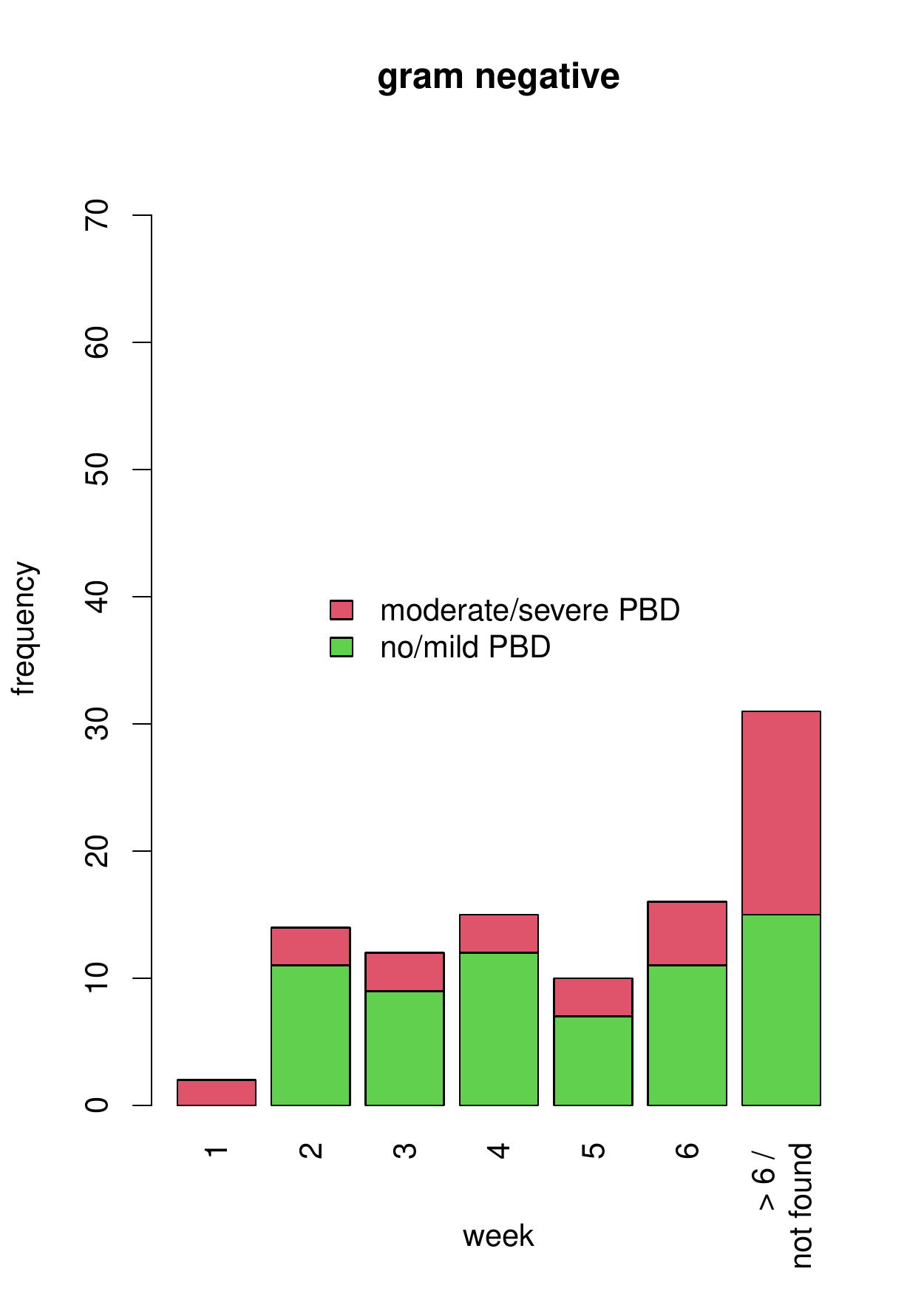}
	\includegraphics[width=57mm]{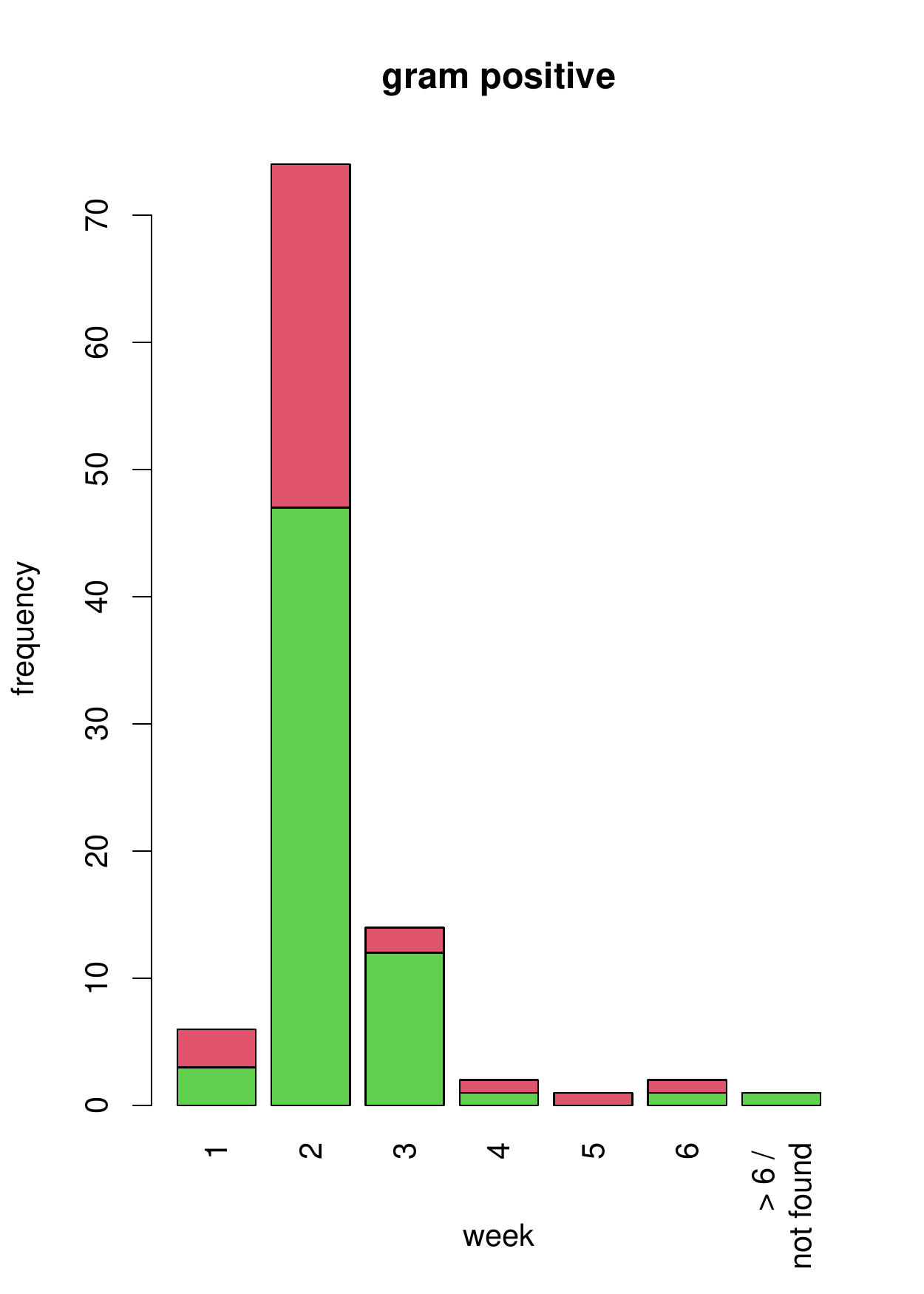}
	\includegraphics[width=57mm]{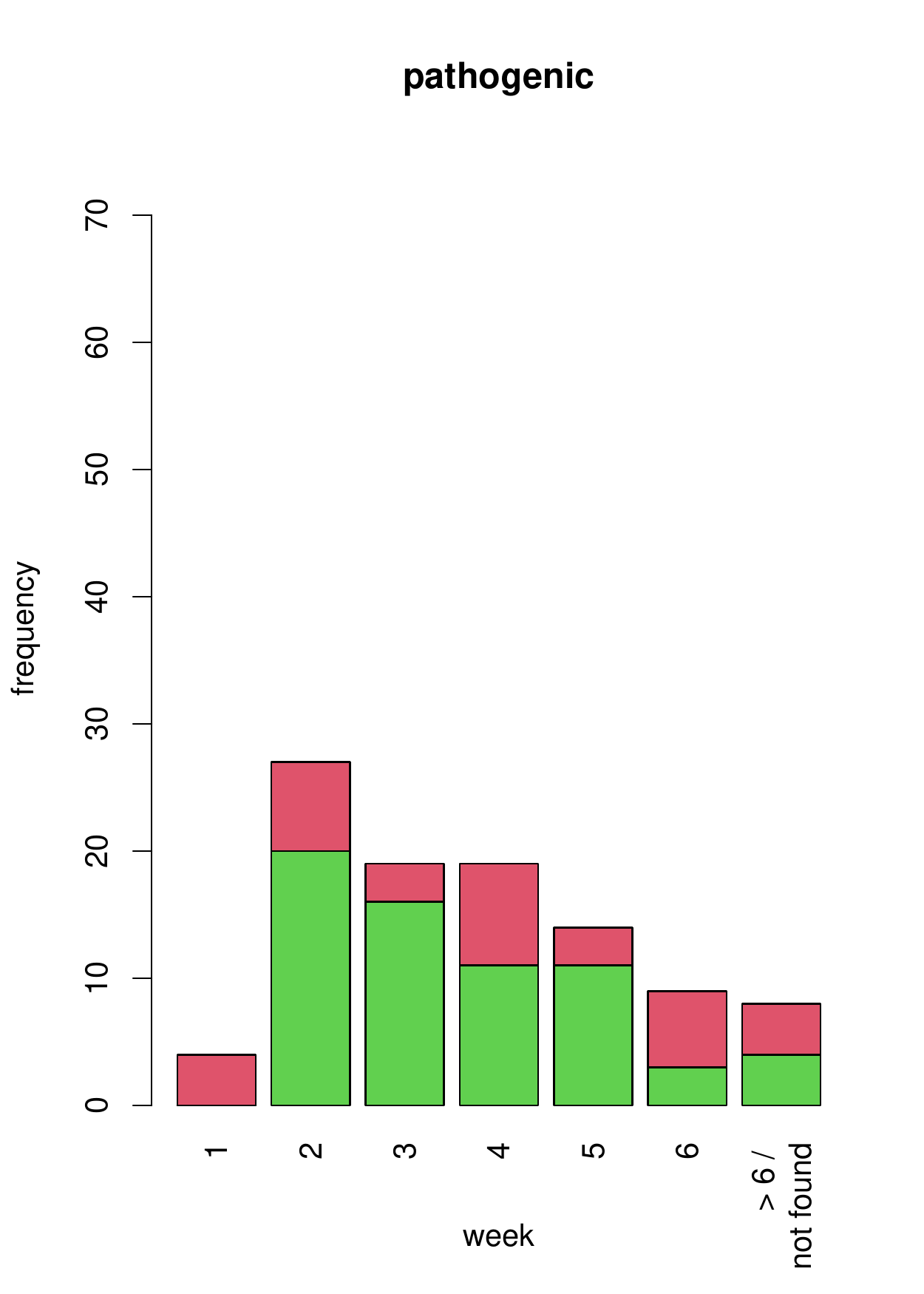}
	\caption{\label{fig:data} Detection of oral bacteria and occurrence of BPD.}
\end{figure}

Bronchopulmonary Dysplasia (BPD) is a chronic lung disease often found in preterm infants with lungs not fully developed. Disturbance of lung development and severity of BPD is caused by various peri- and postnatal factors including prematurity itself, as well as pre- and postnatal infections~\cite{GroEtal2018}. BPD is measured on ordinal scale with grades 0, 1, 2, 3, but often dichotomized as 0:~`no/mild BPD' and 1:~`moderate/severe BPD'. One goal of the study reported here is to investigate whether the time after birth some specific bacteria were found for the first time in the children's upper airway has an effect on BPD. The study was conducted following the rules of the Declaration of Helsinki of 1975, revised in 2013. The retrospective analysis was approved by the ethics committee of the Justus-Liebig-University Giessen (Az 97/14). Initially, $n = 102$ preterm infants with a birth weight $<$1000~g and gestational age $\le$ 32 $+$ 0 weeks were analyzed within a retrospective cohort study at the tertiary perinatal center of Justus-Liebig-University Giessen (Germany) between January 2014 and June 2017. Two infants, however, had to be excluded from further analyses at some point due to missing information on some bacterial colonization. Earlier analyses already showed that the later bacteria were found, the lower the risk of developing BPD~\cite{LauEtal2020}. However, it is not fully understood yet which specific bacteria have an effect, and in which way. Figure~\ref{fig:data} shows occurrence of BPD depending on the week (after birth) three types of bacteria (gram negative/positive and pathogen) were found for the first time in the upper airway of the respective child. Although `time' is supposed to be a continuous variable, information is only available in a discretized way here, because samples were only obtained once a week. Furthermore, the last category is open/censored. If an observation is falling in the last category, we only know that until week six the respective germ had not been found yet. So the covariate `time/week' may only be considered as discrete but ordinal.

Besides the information on bacteria, some additional risk factors need to be taken into account, such as the weight and sex of the child, the number of days antibiotics and steroids were given, or information on multiples. For doing so, a logit model with categorical/ordinal predictor `week' and additional, potentially confounding covariates may be fit. Typically, a categorical covariate is included as dummy-coded factor, ignoring, however, the information on the categories ordering (if so). In the case presented, additional problems are caused by the fact that some categories only have a few observations (see Figure~\ref{fig:data}), and sometimes all those observations are falling in the same response category. The latter means that in a logit model fitted via maximum likelihood corresponding regression coefficients tend towards $\pm \infty$.  

For preventing numerical problems like inflating regression coefficients, penalization can be a viable solution~\cite{HoeKen1970}. Furthermore, penalty terms can be used for exploiting/respecting the covariates' ordinal scale level. Following Gertheiss \& Tutz~\cite{GerTut2009, TutGer2014, TutGer2016}, for instance, a difference penalty might be put on adjacent dummy coefficients of the ordinal factor when fitting the model; an approach that has already been applied successfully in medical research, see, e.g., \cite{CieEtal2014, GlaRos2015}. In our application, however, the question naturally arises how to test for significance of the ordinal predictor in the penalized setting. In a linear model with normal errors, this can be done using a (restricted) likelihood ratio test~\cite{CraRup2004, CraEtal2005, GreEtal2008, SchGreKue2008}, after rewriting the ordinal penalty as a mixed model~\cite{GerOeh2011, Ger2014} (see also Section~\ref{Sec:OGAM}). However, the corresponding test is not available for generalized linear models, such as the logit model considered here.  

In this study, we will discuss how technology developed for generalized additive models~\cite{HasTib1990, Wood2017} can be used to fit generalized linear and additive models with ordinal smoothing penalty, and conduct further statistical inference. In more detail, the rest of the paper is organized as follows. In Section~\ref{Sec:OGAM}, we will revisit generalized additive models, the ordinal penalty, and corresponding statistical inference. In Section~\ref{Sec:NumExp}, we will carry out some numerical experiments to investigate size and power of the test(s) described in Section~\ref{Sec:OGAM} as well as coverage of confidence intervals. Furthermore, we will compare accuracy of penalized estimates to well established standard procedures typically used for data analysis with ordinal covariates. The BPD data will be analyzed in more detail in Section~\ref{Sec:BPD}, and Section~\ref{Sec:Conclude} concludes.

\section{Ordinal Predictors in Generalized Additive Models}\label{Sec:OGAM}

Given a response $y$ with distribution from a simple exponential family, and a set of covariates $x_1,\ldots,x_p$, a generalized additive model~\cite{HasTib1990} has the form: 

\begin{equation}\label{eq:gam}
\eta = \alpha + f_1(x_1) + \ldots + f_p(x_p), \;\; \mu = h(\eta),
\end{equation}
where $\mu$ is the (conditional) mean of $y$ given the covariates, $h$ is a (known) response function, and $\eta$ is comparable to the linear predictor in generalized linear models~\cite{NelWed1972, McCNel1989}. The difference to a generalized linear model is that non-linear functions $f_j$, $j=1,\ldots,p$, are allowed in $\eta$, but still the structure of $\eta$ is additive. Of course, if $f_j$ are restricted to be linear, a generalized linear model is obtained as a special case. In a (generalized) additive model, however, it is usually only assumed that functions $f_j$ are reasonably smooth; and one way to fit such models, as for instance implemented in the popular R package \texttt{mgcv}~\cite{RCore2020, Wood2017}, is to specify a set of basis functions for each predictor and to employ an appropriate, quadratic smoothing penalty on the corresponding basis coefficients. More precisely, we assume that
\begin{equation}
f_j(x) = \sum_{r=1}^{q_j} \beta_{jr} B_{jr}(x),
\end{equation}
with $B_{j1}(x),\ldots,B_{jq_j}(x)$ being the set of basis functions chosen for function $f_j$, and $\beta_{j1},\ldots,\beta_{jq_j}$ the corresponding basis coefficients. So by fitting the basis coefficients, the function $f_j$ is fitted implicitly. For instance, a popular choice in case of a continuous $x$ is the so-called B-spline basis, leading to $f_j$ being modeled as a spline function; see, e.g., de Boor~\cite{deBoor1978}, Dierckx~\cite{Dierckx1993} or Eilers \& Marx~\cite{EilMar1996} for details on (B-)splines. The big advantage of the basis functions approach is that after plugging-in the data observed $x_{ij}$, $i=1,\ldots,n$, $j=1,\ldots,p$, vector $f_j = (f_j(x_{1j}),\ldots,f_j(x_{nj}))^\top$ can be written as $B_j\beta_j$, with matrix $(B_j)_{ir} = B_{jr}(x_{ij})$ and vector $\beta_j = (\beta_{j1},\ldots,\beta_{jq_j})^\top$. So model (\ref{eq:gam}) can be written as a (generalized) linear model with coefficients $\beta_{jr}$, $j=1,\ldots,p$, $r=1,\ldots,q_j$, and basis coefficients can be fitted accordingly, in particular by maximum likelihood/Fisher scoring. 

The problem with this approach is that typically a rather large number of basis functions needs to be chosen for being sufficiently flexible with respect to the type of functions that can be fitted. With a large basis, however, the number of basis coefficients to be fitted becomes large, too. As a consequence, resulting functions tend to be wiggly and thus hard to interpret. Therefore, a penalty term $J_j(\beta_j)$ is typically added for each covariate $x_j$, penalizing wiggly basis coefficients and thus wiggly functions $f_j$. When combining the B-spline basis with a quadratic difference penalty on adjacent $\beta_{jr}$, for instance, we obtain the so-called P-spline~\cite{EilMar1996, MarEil1998, EilMar2002}.\\

Now suppose you have a categorical predictor $x_j$ with levels $1,\ldots,k_j$. Then there is a somewhat natural basis: the basis of (dummy) functions ($l=1,\ldots,k_j$)
\begin{equation}\label{eq:obasis}
B_{jl}(x_j) = \left\{\begin{array}{rr}
1 & \mbox{if } x_j = l,\\
0 & \mbox{otherwise}.
\end{array}
\right.
\end{equation}
Since we know that $x_j$ can only take values $1,\ldots,k_j$, we do not need to think about the type and number of basis functions, placing of knots, etc., as we usually do with continuous covariates. For means of identifiability, however, some linear restrictions need to placed on the basis/dummy coefficients $\beta_{jl}$. Typically, this is done by specifying a so-called reference category, e.g., category $1$ for categorical predictor $x_j$, and setting the corresponding $\beta_{j1} = 0$. However, one may also set $\sum_l \beta_{jl} = 0$. In case of a continuous $x_j$ a popular constraint is
\begin{equation}\label{eq:constr}
\sum_{i=1}^n f_j(x_{ij}) = 0 \; \mbox{ for all } j,
\end{equation}
which translates into $\sum_{l} n_{jl}\beta_{jl} = 0$ for categorical predictor $x_j$, with $n_{jl}$ being the number of samples with level $l$ being observed for $x_j$. Since the latter is the typical constraint in generalized additive models, also used in \texttt{mgcv}, we will use (\ref{eq:constr}) here as well (if not stated otherwise). After having fit the functions/basis coefficients, however, one can switch between constraints easily, because switching between the constraints mentioned above for some $f_j$ yields an equivalent model as it simply means a vertical shift of the entire function $f_j$ and a corresponding change in the constant $\alpha$ in~(\ref{eq:gam}). 

If now a (quadratic) difference penalty is put on the basis/dummy coefficients, this gives exactly the smoothing penalty as mentioned above~\cite{GerTut2009, TutGer2014, TutGer2016}. More precisely, with $\beta_{jl}$, $l=1,\ldots,k_j$, denoting the dummy coefficient of level $l$, the penalty primarily used is the first-order penalty
\begin{equation}\label{eq:pen1}
J_j(\beta_j) = \sum_{l=2}^{k_j} (\beta_{jl} - \beta_{j,l-1})^2,
\end{equation}
with $\beta_j =  (\beta_{j1},\ldots,\beta_{jk_j})^\top$. Alternatively, however, the second-order penalty
\begin{equation}\label{eq:pen2}
J_j(\beta_j) = \sum_{l=2}^{k_j-1} (\beta_{j,l+1} - 2\beta_{jl} + \beta_{j,l-1})^2,
\end{equation} 
penalizing deviations from linearity can be used as well~\cite{GerOeh2011}. The strength of the penalty put on $\beta_j$ is determined by a parameter $\lambda_j$, which may be different for different predictors $x_j$. Then, parameters can be fitted by maximizing the penalized log-likelihood
\[
l_p(\beta) = l(\beta) - \sum_{j=1}^p \lambda_j J_j(\beta_j),
\]
where $\beta$ denotes the vector collecting all basis coefficients and, if so, additional regression coefficients of covariates included parametrically, i.e., in terms of linear effects; $l(\beta)$ is the usual, unpenalized log-likelihood. In case of $\lambda_j = 0$ for all $j$, the usual maximum likelihood estimates are obtained. If $\lambda_j \rightarrow \infty$ in case of penalty (\ref{eq:pen1}), fitted $f_j(x) = 0$ for all $x \in \{1,\ldots,k_j\}$ is obtained because of the constraint (no matter which one is chosen from the options given above). If penalty~(\ref{eq:pen2}) is chosen, large $\lambda_j$ leads to a function $f_j$ being linear in the class labels $1,\ldots,k_j$.

One of the benefits of considering ordinal predictors along with penalties (\ref{eq:pen1}) and (\ref{eq:pen2}) in the framework of generalized additive models is that after implementing basis~(\ref{eq:obasis}) in the appropriate way, \texttt{gam()} from \texttt{mgcv} can be used directly to fit a generalized linear/additive model with ordinal predictor(s) as needed for the BPD data. Besides pure model fitting, however, this provides us with additional tools; in particular, built-in estimation of the penalty/smoothing parameter(s), further statistical inference, such as confidence intervals, and checking significance of smooth terms. A prerequisite for at least some of those tools, however, is to rewrite our model with quadratic smoothing penalty as a (generalized) linear mixed model. 

\subsection{Mixed Model/Bayesian Interpretation and Estimation of Smoothing Parameters}
Starting with penalty~(\ref{eq:pen1}), we may rewrite our model in terms of $\tilde{{\beta}}_j = \tilde{{D}}_1 {\beta}_j$ with
\[ \tilde{{D}}_1 = \left[\begin{array}{cccc}
1 & 0 & \ldots & 0\\
\multicolumn{4}{c}{{D}_1}
\end{array} \right], \mbox{ and } {D}_1 = \left(\begin{array}{cccccc}
-1 & 1 & 0 & 0 & \ldots \\
0 & -1 & 1 & 0 & \ldots \\
\vdots & \ddots & \ddots & \ddots & \ddots\\
\end{array}\right).\]
Let's now denote the sub-vector consisting of the last $k_j-1$ elements of $\tilde{{\beta}}_j$ by ${u}_j = (u_{j1},\ldots,u_{j,k_j-1})^\top$, such that $u_{jl} = \beta_{j,l+1} - \beta_{jl}$ and ${u}_j = {D}_1 {\beta}_j$, with ${D}_1$ from above. The entries of ${u}_j$ are now interpreted as iid random effects with $u_{jl} \sim N(0,\tau_j^2)$~\cite{GerOeh2011, Ger2014, SweCraGer2016}. Then, for given variance parameters $\tau_j^2$, $j=1,\ldots,p$ (note, the random effects' variance may vary between covariates), maximizing the log-likelihood over $\tilde{{\beta}}_j$ yields estimates that are equivalent to the smoothed dummies obtained via penalty~(\ref{eq:pen1}), with a one-to-one correspondence of penalty parameter $\lambda_j$ and variance parameter $\tau_j$. Alternatively, smoothed dummy coefficients can be derived in a Bayesian framework (as the mode of the posterior density) by putting a Gaussian random walk prior (with prior variance $\tau_j^2$) on the dummy coefficients~\cite{GerTut2009}. Analogously, penalty~(\ref{eq:pen2}) can be derived/interpreted in a mixed model/Bayesian framework. For that purpose, we replace $\tilde{{D}}_1$ above by $\tilde{{D}}_2$ with
\[ \tilde{{D}}_2 = \left[\begin{array}{ccccc}
1 & 0 & 0 & \ldots & 0\\
0 & 1 & 0 & \ldots & 0\\
\multicolumn{5}{c}{{D}_2}
\end{array} \right], \mbox{ and } {D}_2 = \left(\begin{array}{ccccccc}
1 & -2 & 1 & 0 & 0 & \ldots \\
0 & 1 & -2 & 1 & 0 & \ldots \\
\vdots & \ddots & \ddots & \ddots & \ddots & \ddots\\
\end{array}\right).\]
Then, the last $k_j-2$ elements of $\tilde{{\beta}}_j$ are denoted by ${u}_j = (u_{j1},\ldots,u_{j,k_j-2})^\top$, such that $u_{jl} = \beta_{j,l+2} - 2\beta_{j,l+1} + \beta_{jl}$, and as before $u_{jl} \sim N(0,\tau_j^2)$ are interpreted as iid random effects~\cite{GerOeh2011, Wood2017}.

Both approaches, the mixed model and Bayesian interpretation, can be used for determining the variance components $\tau_j^2$, and thereby the penalty parameters $\lambda_j$. In theory, we may integrate out the random effects from the joint density of the response and random effects, giving the marginal likelihood of the fixed effects and the variance parameters. Maximizing this likelihood leads to ML estimates of the fixed effects and variance parameters. In generalized linear mixed models, however, calculating the integral analytically is often not possible, and also numerically demanding. The standard approach is so-called \emph{Laplace approximation}~\cite{BreCla1993}, which essentially cycles through the penalized log-likelihood given above and plugging-in the corresponding estimates of regression/basis coefficients to obtain an approximate profile likelihood for the variance parameters that can be maximized. In the Bayesian framework, the smoothed dummies are then interpreted as an \emph{empirical} Bayes estimator, since $\tau_j^2$ are estimated from the data. In a fully Bayesian approach, we could choose a hyper-prior, e.g., an Inverse Gamma, for $\tau_j^2$ and apply Markov Chain Monte Carlo (MCMC); but we will focus on the mixed model/empirical Bayes approach here.

A problem with ML estimation of variance parameters is that those estimates are typically biased downwards, that is, the true variance tends to be underestimated, in particular if the number of fixed effects is large. As an alternative to ML which reduces this bias, so-called \emph{restricted} maximum likelihood (REML) estimation has been proposed, which can be motivated in different ways~\cite{PatTho1971, Har1974, Har1977, LaiWar1982, FahEtal2013}. Eventually, in the linear mixed model, the (profile) log-likelihood is (additively) corrected such that the number/structure of fixed effects is taken into account. In generalized mixed models, this can be done analogously within Laplace approximation. It should be noted that REML cannot be used to compare models with different fixed effect structure. Nevertheless, REML is very popular for estimating variance components in mixed models, due to the reduced bias, and hence for determining smoothing/penalty parameters in (generalized) additive models as well~\cite{Wood2011}. Besides likelihood-based methods, also prediction error methods such as (generalized) cross-validation~\cite{Wood2008} or the AIC~\cite{WooPyaSae2016} can be used for smoothness selection.

\subsection{Confidence Intervals and Significance of Smoothed Dummies}
In particular the Bayesian interpretation of quadratic smoothing penalties is useful to derive confidence intervals. In the Gaussian identity link/linear mixed model case, we can derive the covariance matrix of the regression/basis coefficients' posterior distribution. In the generalized case the corresponding matrix from the penalized iteratively weighted least squares algorithm (PIRLS) used to estimate the parameters is taken. Using this matrix, let's say $V_{\beta}$, we can calculate (point-wise) credible intervals for function $\hat{f}_j = B_j\hat{\beta}$, denoting the vector of $\hat{f}_j(x)$ values at evaluation points $x_{ij}$. For each element $\hat{f}_{ji}$ of $\hat{f}_j$, we obtain an approximate $(1-\alpha)100$\% credible interval via $\hat{f}_{ji} \pm z_{1-\alpha/2}\sqrt{v_{ji}}$, where $v_{ji}$ is the $i$th diagonal element of $B_jV_{\beta}B_j^\top$, and $z_q$ is the $q$-quantile of the standard normal distribution. It turned out that those credible regions also have good frequentist coverage rates in generalized additive models with continuous covariates~\cite{Nychka1988, MarWoo2012}. In Section~\ref{Sec:NumExp}, we will investigate whether this is also the case with smoothed ordinal effects.

Closely related to the confidence intervals described above is testing significance of smooth terms. Specifically, we would like to test null hypothesis $H_0: f_j(x) = 0$ for all potential values $x$ of predictor $x_j$. With ordinal predictors, this means, for all levels of the predictor of interest. Analogously to the confidence intervals above, let $\hat{f}_j = B_j\hat{\beta}$ denote the vector of $\hat{f}_j(x)$ evaluated at the predictor levels (i.e., with appropriately chosen dummy-matrix $B_j$). Then, following~\cite{Wood2013, Wood2017}, we can define a Wald-type test statistic
\[T = \hat{f}_j^\top V_j^{-} \hat{f}_j,\]
where $V_j^{-}$ is an appropriately chosen pseudo-inverse of $V_j = B_j V_{\beta}B_j^\top$. Under $H_0$, $T$ (approximately) follows a mixture of $\chi^2$-distributions. For the concrete form of this mixture and $V_j^{-}$, see Wood~\cite{Wood2013, Wood2017}.

When taking the mixed models perspective of penalty~(\ref{eq:pen1}), the null hypothesis above can alternatively be written as $H_0: \tau_j^2 = 0$. In the Gaussian/linear mixed model with only one smooth term/variance component a (restricted) likelihood ratio test can be used~\cite{CraRup2004, CraEtal2005, SchGreKue2008}, which also works well with ordinal data~\cite{Ger2014, GerOeh2011, SweCraGer2016}, and provides extensions/approximations in case of multiple ordinal predictors/smooth terms~\cite{Ger2014, GreEtal2008, SchGreKue2008}. To the best of our knowledge, however, no generalization exists beyond models with Gaussian response. Therefore we will focus on the Wald-type test here.

For instance, when looking at \texttt{summary.gam()} with ordinal predictor $x$ denoting the (first) week of overall oral bacteria detection, and correcting for various (known) risk factors of BPD and/or potential confounders, we have\label{model:overall}:

	\begin{verbatim}
	Family: binomial 
	Link function: logit 
	
	Formula:
	y ~ s(weight, bs = "cr") + SGAsym + sex + mult + s(steroid, bs = "cr") + 
	s(anti, bs = "cr") + s(x, bs = "ordinal", m = 2)
	
	Parametric coefficients:
	            Estimate Std. Error z value Pr(>|z|)
	(Intercept)  -2.8377     0.8537  -3.324 0.000888 ***
	SGAsym        2.0615     1.2461   1.654 0.098061 .  
	sex           1.8593     0.7143   2.603 0.009242 ** 
	mult          0.5417     0.4102   1.321 0.186584    
	---    
	Signif. codes:  0 `***' 0.001 `**' 0.01 `*' 0.05 `.' 0.1 ` ' 1
	
	Approximate significance of smooth terms:
	           edf Ref.df Chi.sq  p-value    
	s(weight)    1      1 13.151 0.000287 ***
	s(steroid)   1      1  4.527 0.033364 *  
	s(anti)      1      1  4.675 0.030600 *  
	s(x)         1      1  5.627 0.017682 *  
	---
	Signif. codes:  0 `***' 0.001 `**' 0.01 `*' 0.05 `.' 0.1 ` ' 1
	
	R-sq.(adj) =  0.438   Deviance explained =   42%
	-REML = 30.444  Scale est. = 1         n = 101
	\end{verbatim}

\vspace{3mm}

Here, \texttt{SGAsym} is an indicator `Small for Gestational Age' defined as birth weight, length and head circumference below the 10th percentile according to the percentiles from the German perinatal registry, \texttt{sex} is an indicator for male, and \texttt{mult} denotes multiples. The child's weight at birth is denoted by \texttt{weight}, \texttt{steroid} is the number of days (before delivery) steroids were given (to the mother), and \texttt{anti} the number of days (after birth) antibiotics were given (to the child). We see that all the smooth effects are fitted as linear eventually (as effective degrees of freedom/edf equal 1), and indicted as significant on the 5\% level (at least), which confirms results obtained earlier for this dataset~\cite{LauEtal2020}. In fact, p-values given above are (virtually) identical to the p-values when fitting a standard logit model with linear effects (e.g., using R function \texttt{glm()}, not shown here) indicating the negative effect of \texttt{x}, i.e., the risk of BPD is increased if overall bacterial colonization is detected early. In general, however, it is not clear yet if p-values provided in the generalized additive model are reliable for the ordinal smoothing penalty. So this needs to be investigated before using the ordinal smoothing approach for analyzing the effects of specific types of bacteria.

\section{Numerical Experiments}\label{Sec:NumExp}

For investigating whether the p-values provided by \texttt{summary.gam()} are reliable when using the ordinal basis/smoothing penalty, we will discuss the results of some simulation studies, starting with the BPD data. 

\subsection{Using the BPD Data}\label{Sec:TestBPD}

\begin{figure}[tb]\centering
	\includegraphics[width=57mm]{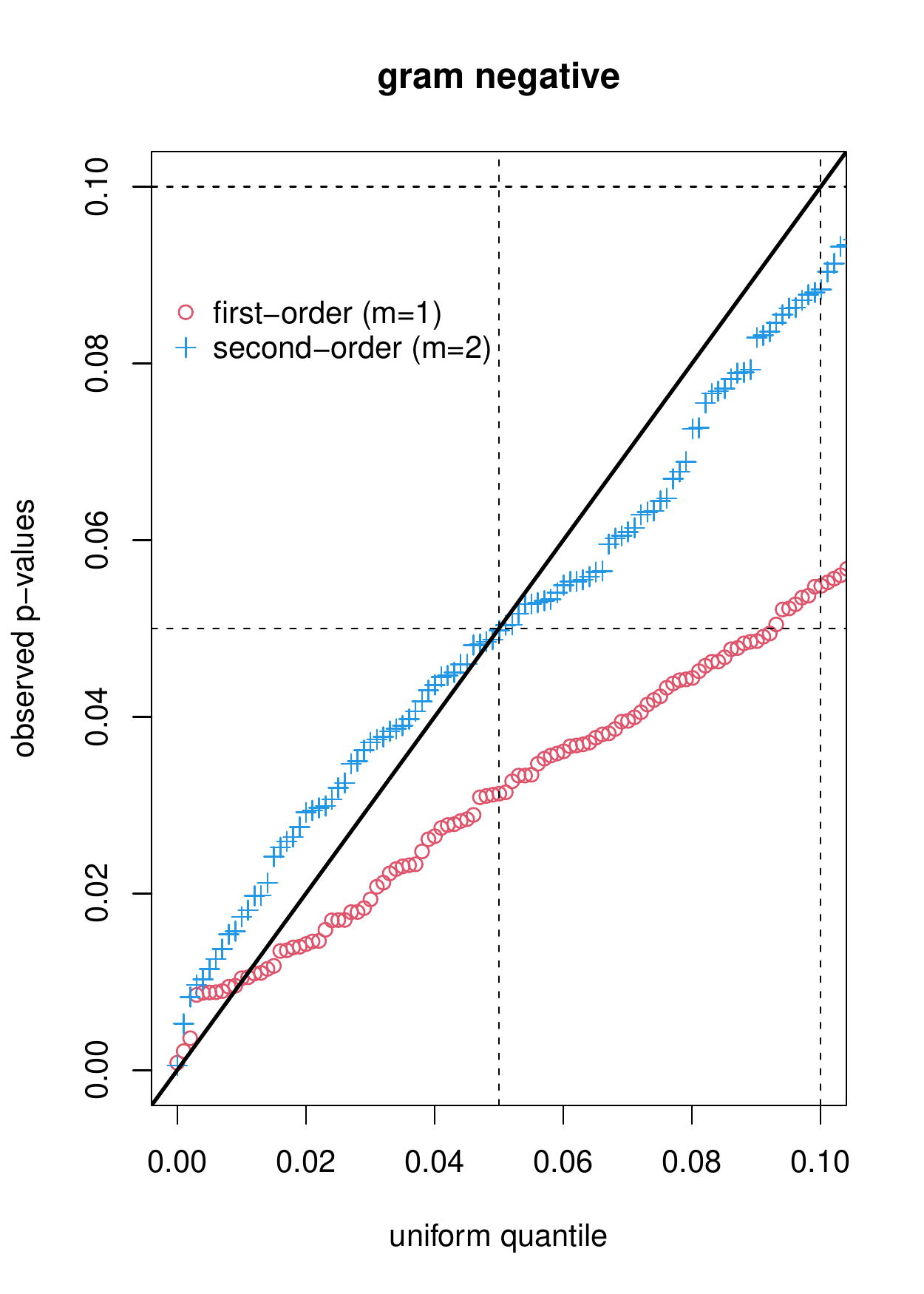}
	\includegraphics[width=57mm]{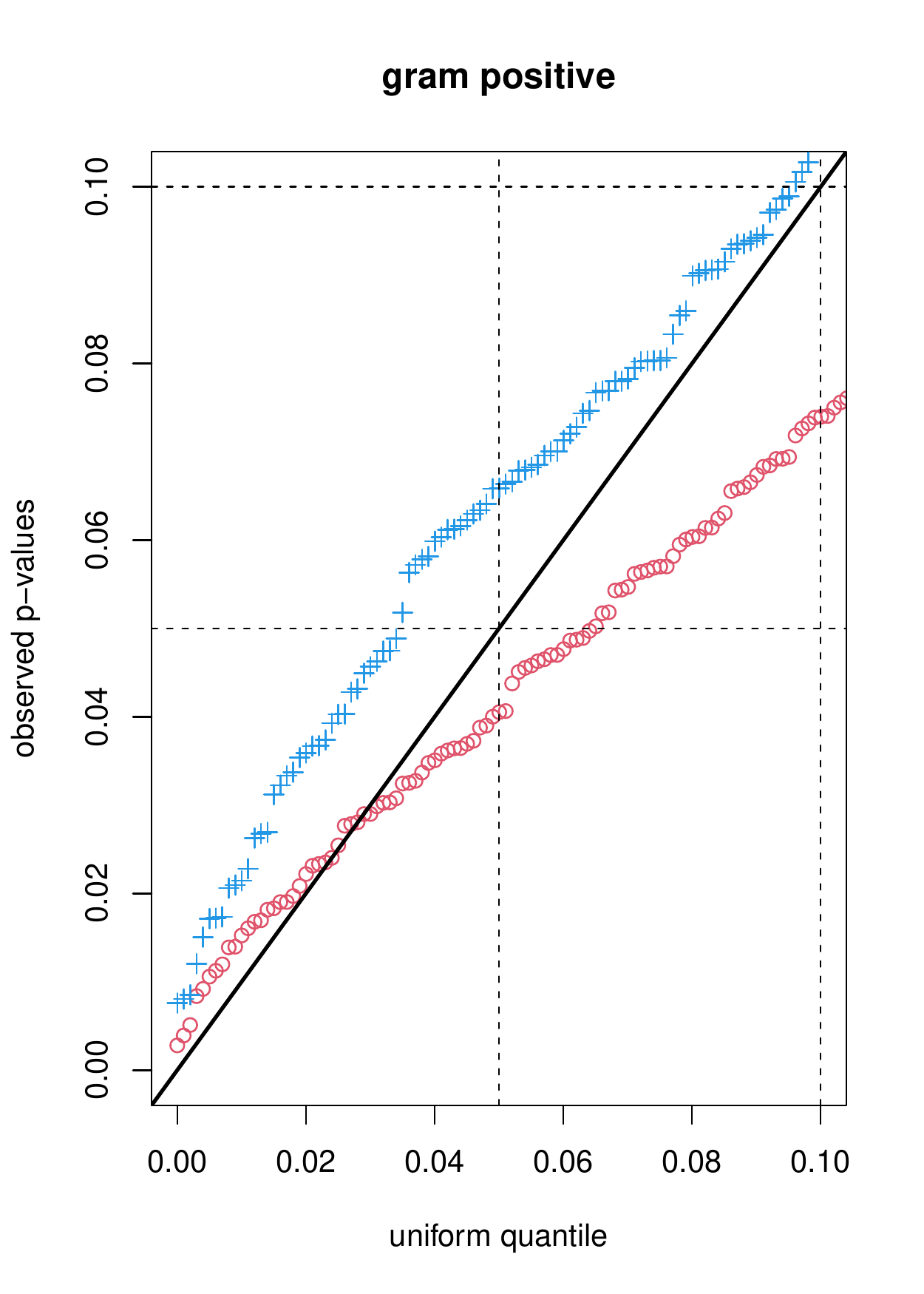}
	\includegraphics[width=57mm]{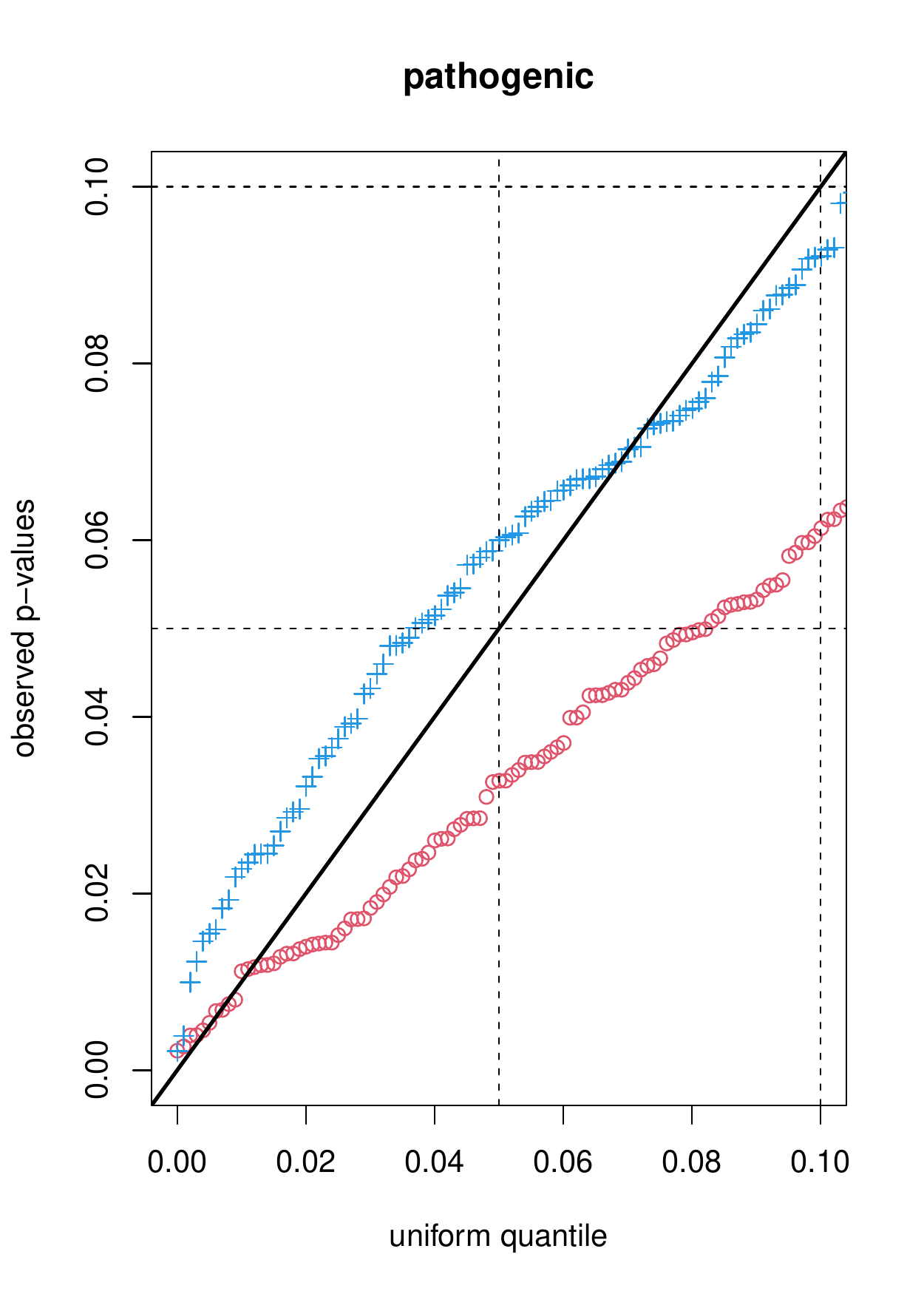}
	
	\vspace{-3mm}
	\caption{\label{gertheiss:fig2} QQ-plots of p-values for first- and second-order penalty (red/blue).}
\end{figure}

We use the confounder model, i.e., the model with information on bacteria removed, to estimate BPD probabilities. That means, the null hypothesis that the effect of (ordinal predictor) $x$ is zero, is true by construction, because fitted BPD probabilities do not depend on $x$, given the other covariates. Using those probabilities, we simulate `new' BPD response data, fit the model with smooth ordinal $x$ added (and smoothing parameter estimated by REML), and report p-values of $x$. For $x$, we use gram negative/positive and pathogenic bacteria, respectively (compare Figure~\ref{fig:data}). For each $x$, this is done 1,000 times; and Figure~\ref{gertheiss:fig2} shows corresponding QQ-plots of p-values observed employing the first- and second-order penalty, respectively. Since the distribution of smaller p-values is particularly relevant when testing with usual $\alpha \le 0.1$, we restrict plotting to that area.  

It is seen that p-values obtained when employing the first-order penalty are typically too small. Problems with the first-order penalty can be explained by the fact that the null space of the corresponding smooth term has dimension zero (compare the \texttt{mgcv} manual). In other words, the null hypothesis in the framework of mixed models (which is used for estimation here), a zero variance component, is on the boundary of the parameter space, which means that standard theory does not apply~\cite{CraRup2004,CraEtal2005}. Results for the second-order penalty, by contrast, look very encouraging. As a consequence of the findings on the BPD data, we will only consider the second-order penalty when further investigating size and power below.

\subsection{Further Studies on Size and Power}\label{Sec:SizePower}

\begin{figure}[t]\centering
	
	\includegraphics[width=56mm, height=79mm]{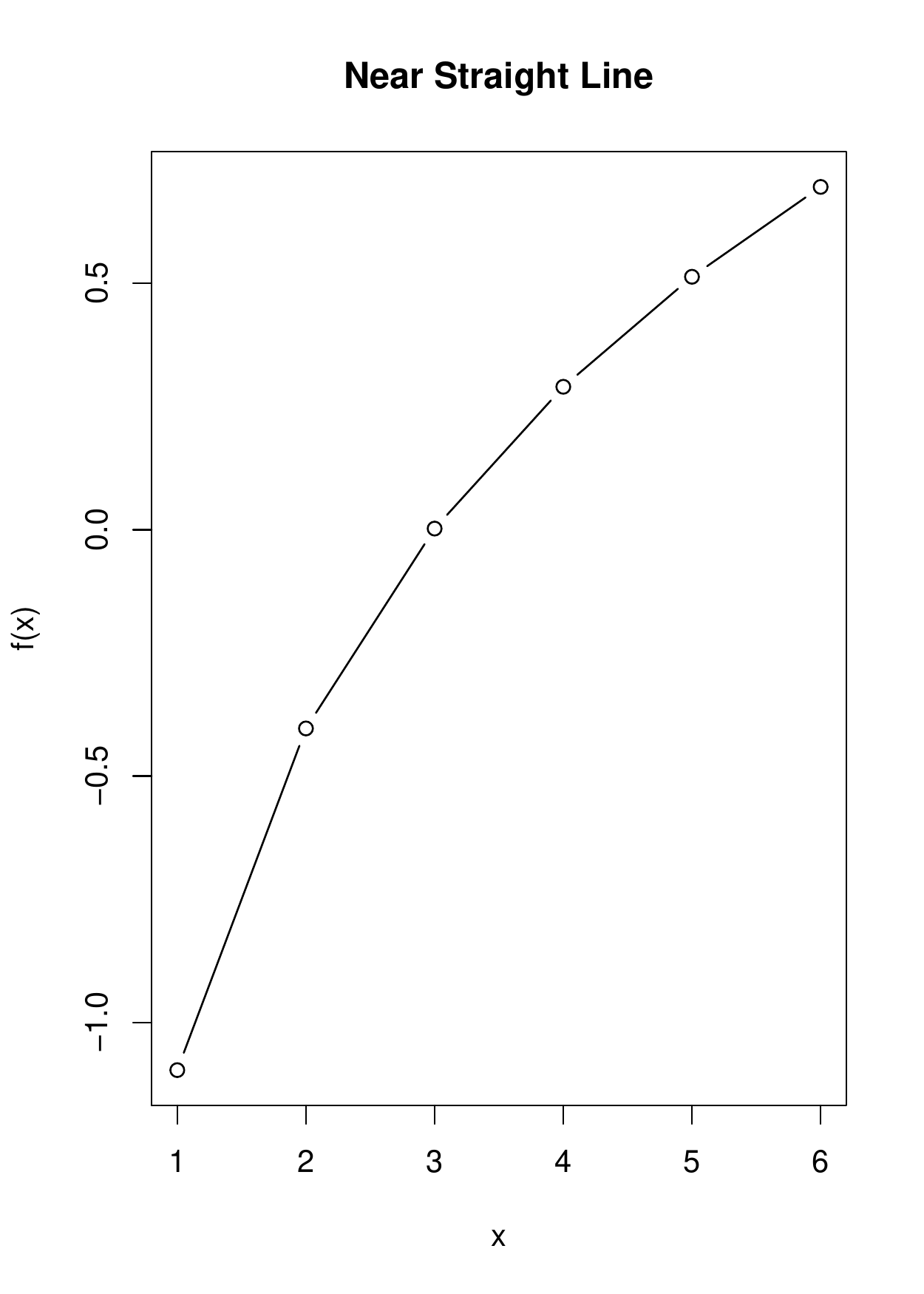}
	\includegraphics[width=56mm, height=79mm]{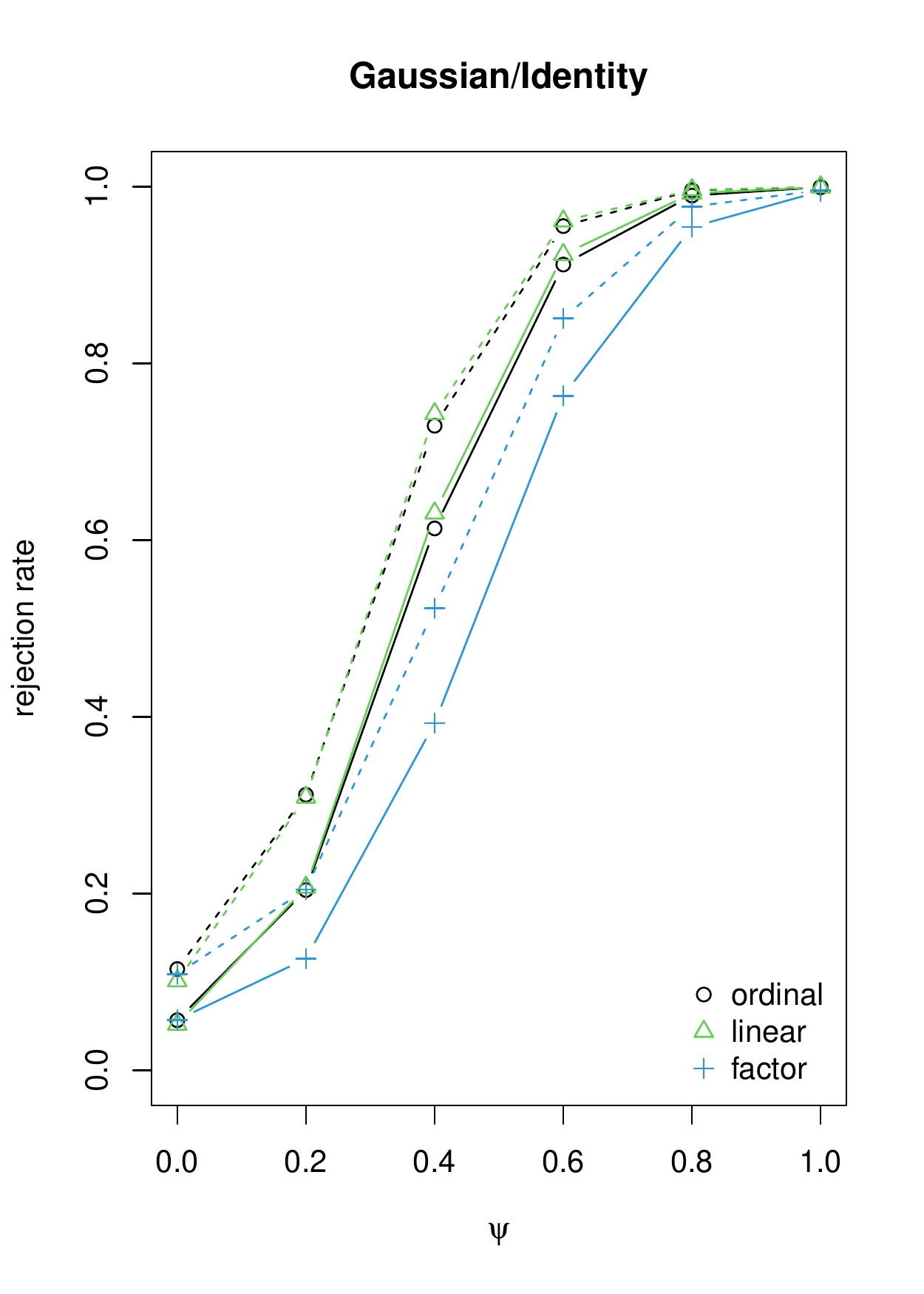}
	\includegraphics[width=56mm, height=79mm]{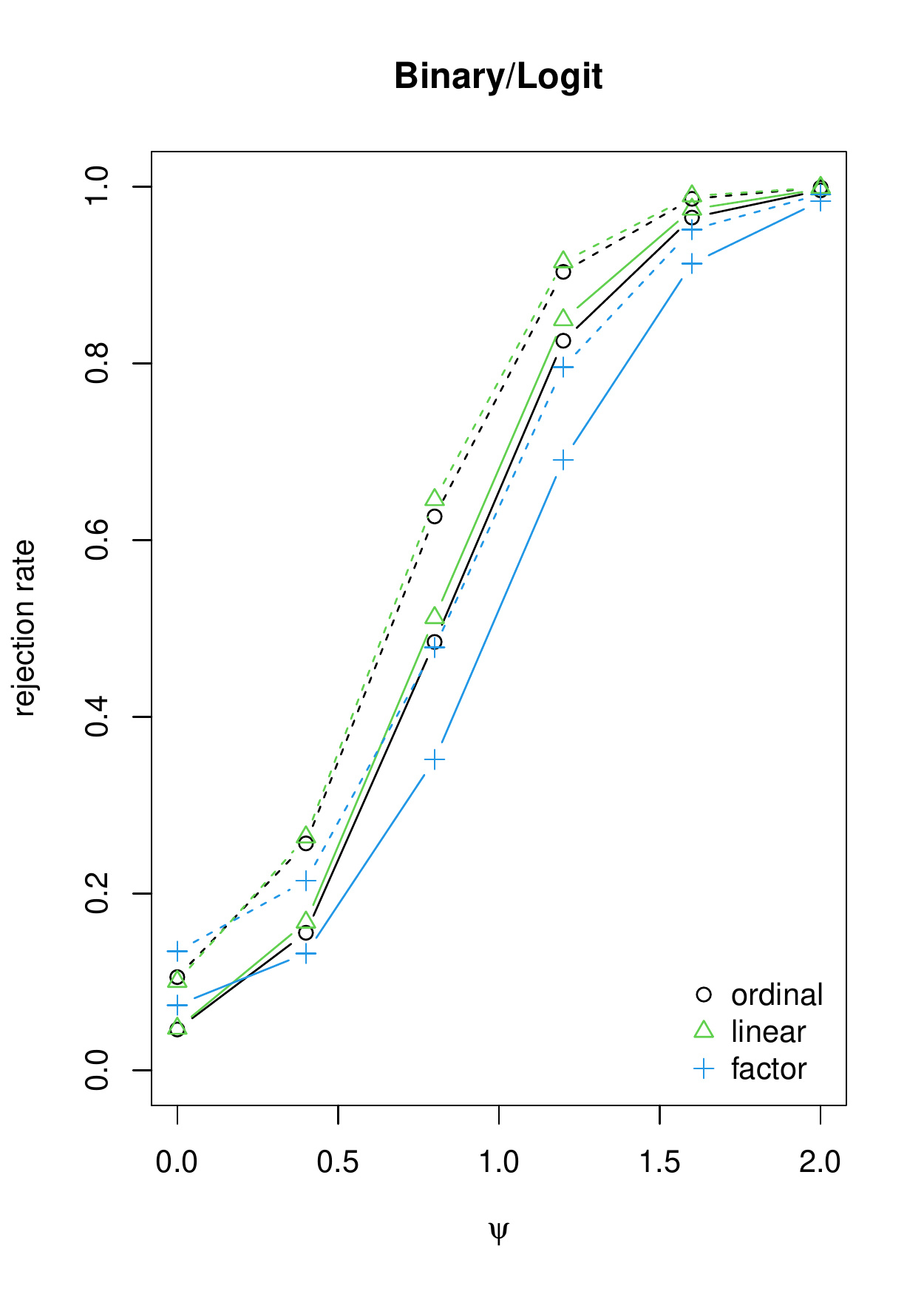}
	
	\vspace{-2mm}
	\includegraphics[width=56mm, height=79mm]{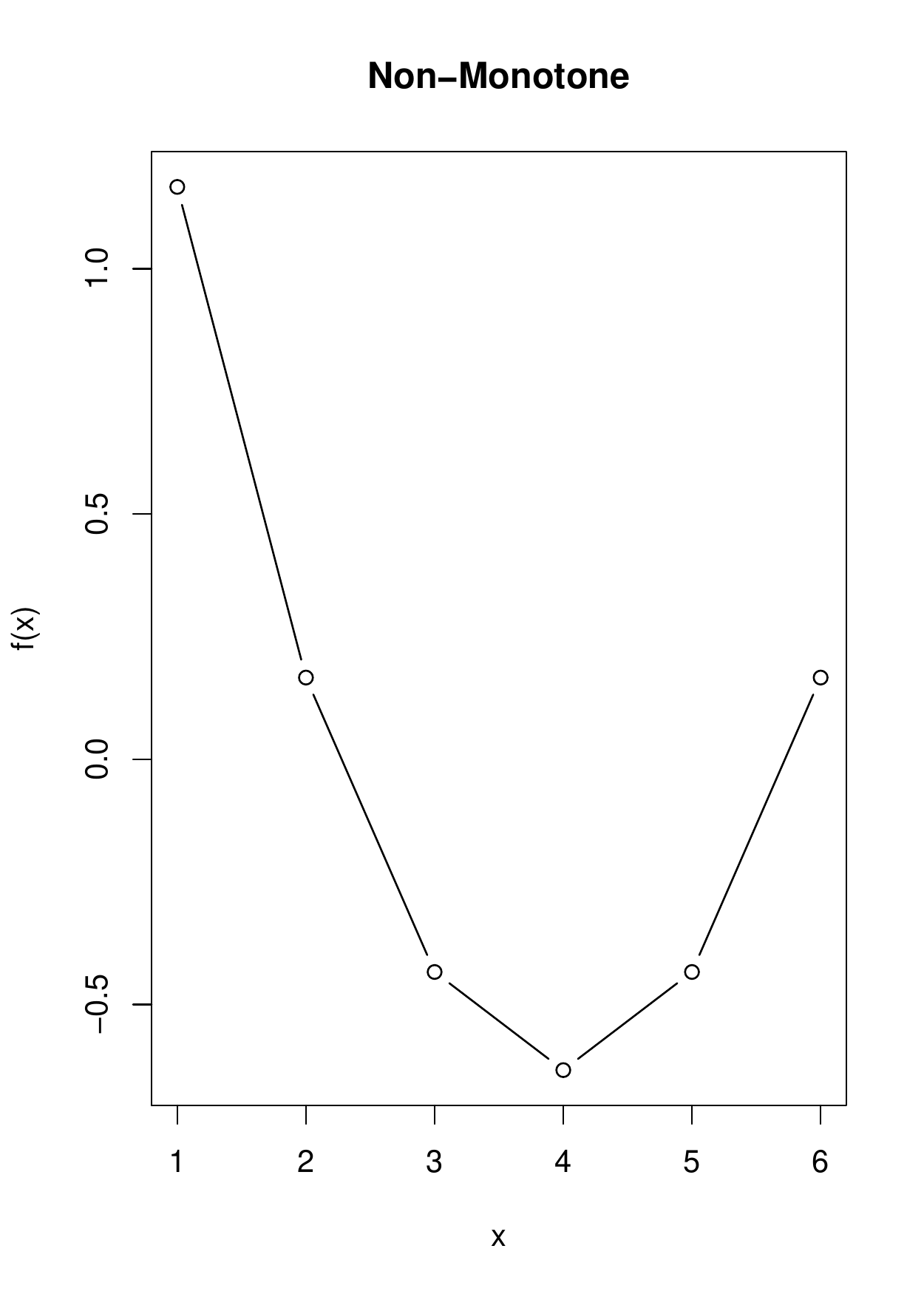}
	\includegraphics[width=56mm, height=79mm]{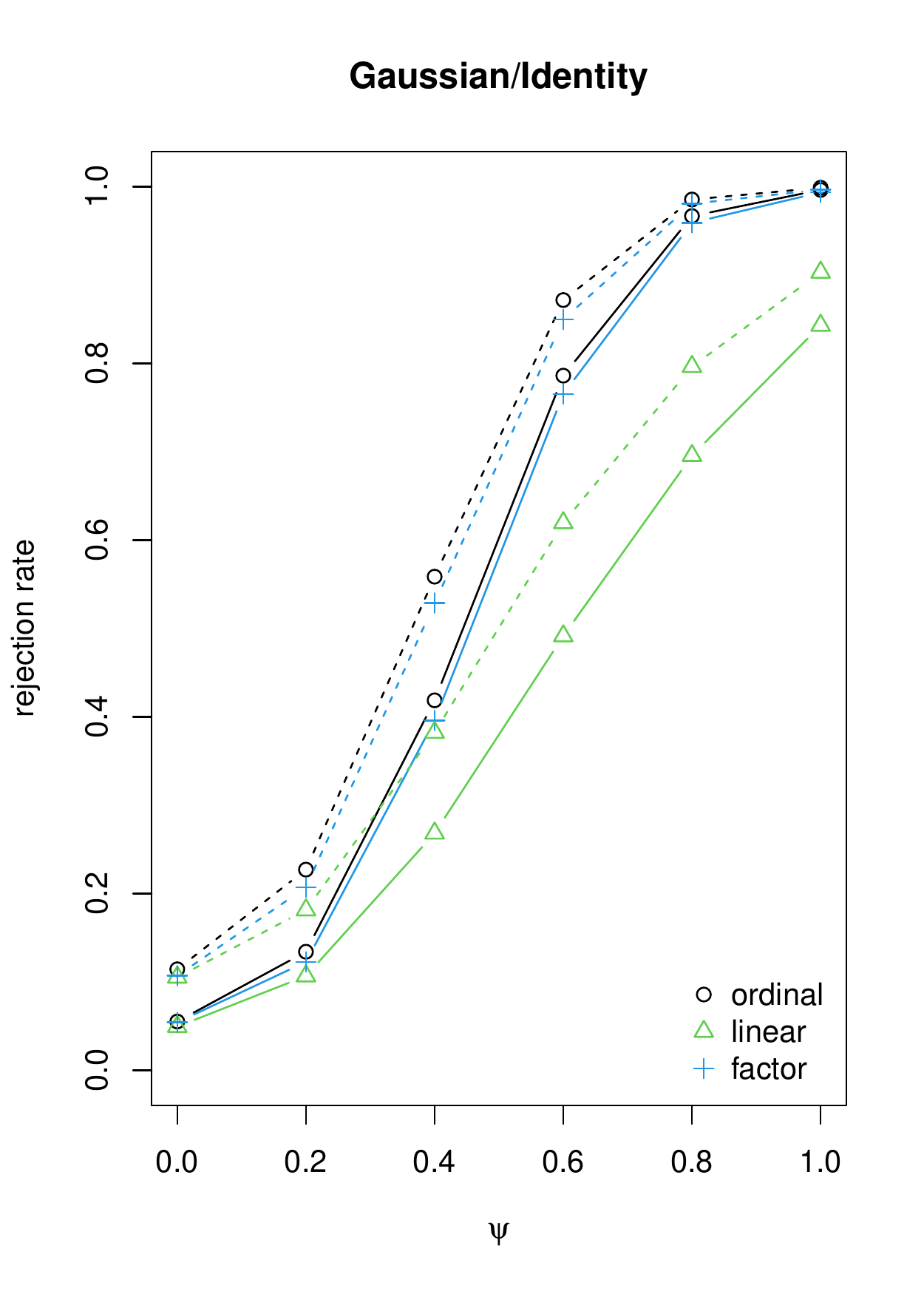}
	\includegraphics[width=56mm, height=79mm]{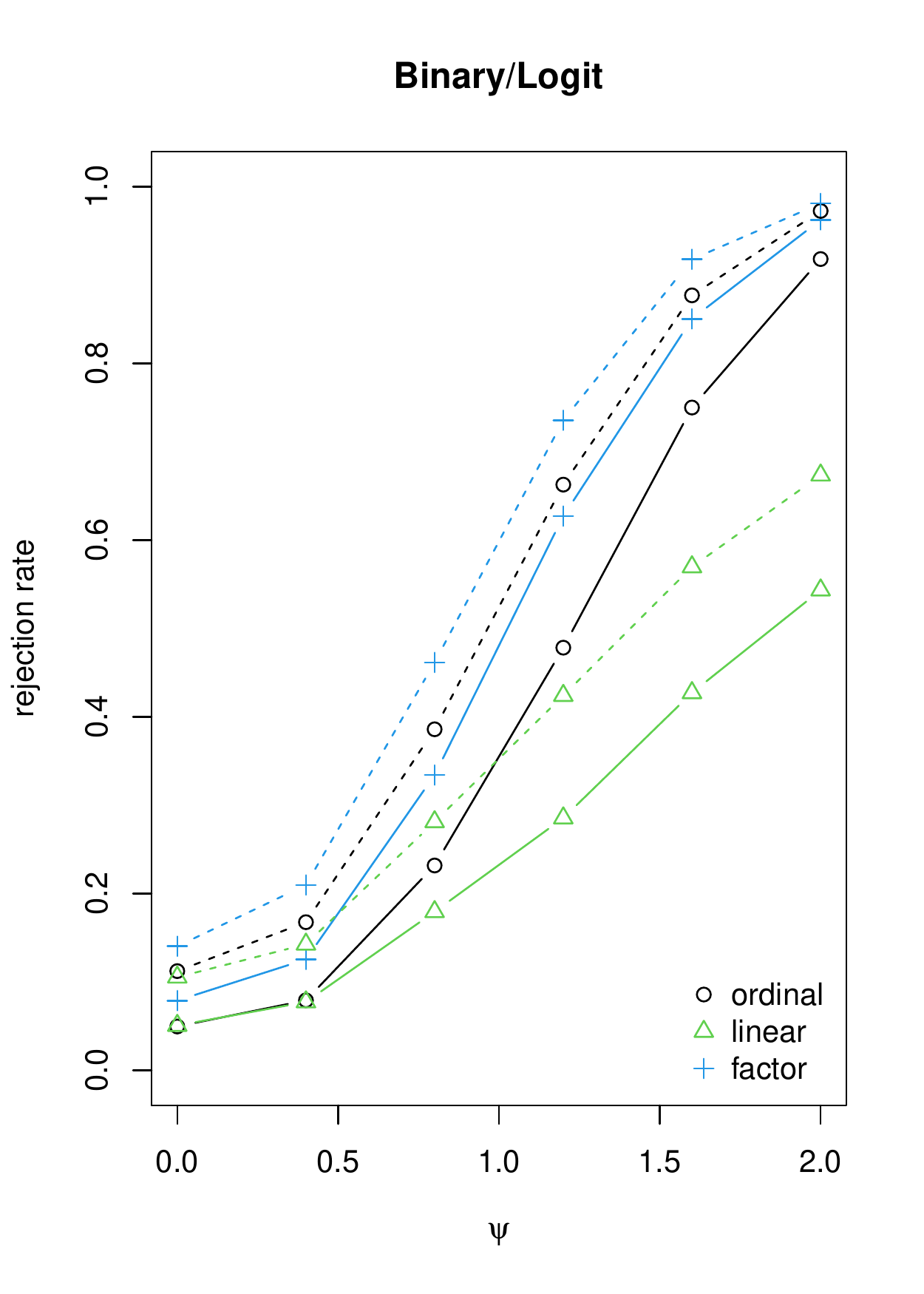}

	\vspace{-3mm}
	\caption{\label{fig:power} Size and power for various tests and settings with $\alpha = 0.05$ (solid) or $\alpha = 0.1$ (dashed); top row refers to near straight line, bottom row to non-monotone scenario.}
\end{figure}

Given a continuous covariate $z \sim U(0,1)$ and an ordinal predictor $x$ uniformly sampled from $\{1,\ldots,6\}$, 
we assume
\begin{itemize}
	\item a Gaussian distribution for $y$ with $\mu = 3\sqrt{z} + \psi f(x)$, and standard normal errors,
	\item a logit model with linear predictor $\eta = -2 + 4\sqrt{z} + \psi f(x)$.
\end{itemize}
In both cases, parameter $\psi \ge 0$ determines the effect size of $x$. For~$f$, we consider the two functions given in Figure~\ref{fig:power} (left). Furthermore size and power over 10,000 replicates with sample size 100 each are shown for the Gaussian/identity (center) and binary/logit case (right), $\alpha=0.05$ (solid) and $\alpha=0.1$ (dashed), and three different tests: besides the smooth effect (black), we consider including $x$ as parametric/linear term (green) or nominal factor (blue). It is seen that the ordinal smoothing penalty adapts well to the situations considered. On the one hand, if the true, underlying function is close to linear (center), the corresponding test behaves almost identically to the parametric/linear specification. On the other hand, in the very non-linear/non-monotone case, its power is similar to the unpenalized nominal/factor modeling. Although the latter seems to be better in the non-monotone, binary/logit case (bottom right), it also appears that the analysis of deviance/GLRT used here produces a rejection rate for effect size zero that is somewhat too high in the logit model (i.e., p-values are too small). In summary, our studies indicate that the ordinal penalty works well in generalized additive models (with smoothing parameters estimated by REML), and testing as provided by \texttt{summary.gam()} in \texttt{mgcv} is safe for the second-order penalty, with power highly competitive to both linear and ordinary factor modeling.

\subsection{Confidence Intervals \& Estimation Accuracy}

\begin{figure}[tb]\centering
	\includegraphics[width=45mm]{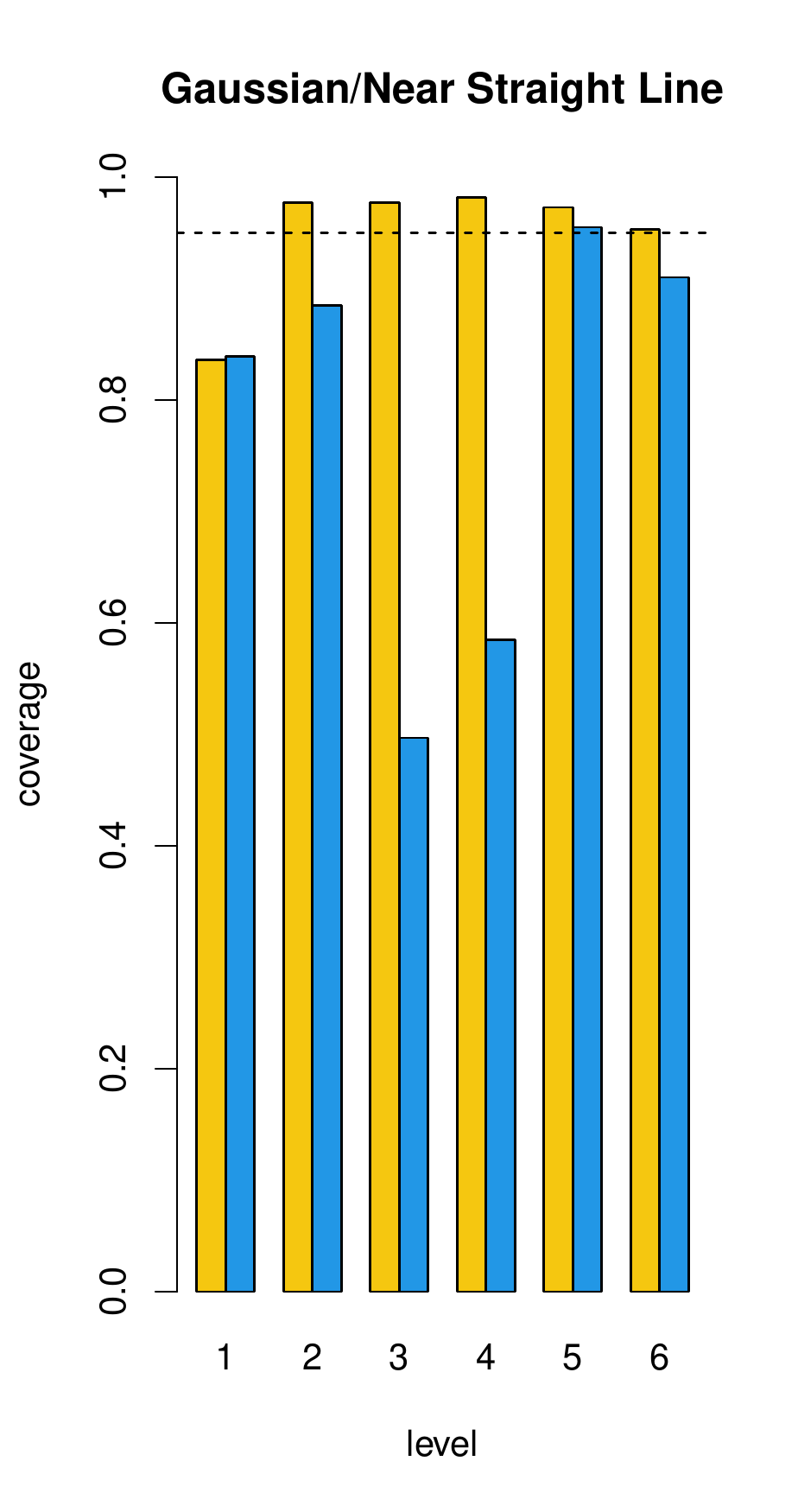}
	\hspace{-4mm}
	\includegraphics[width=45mm]{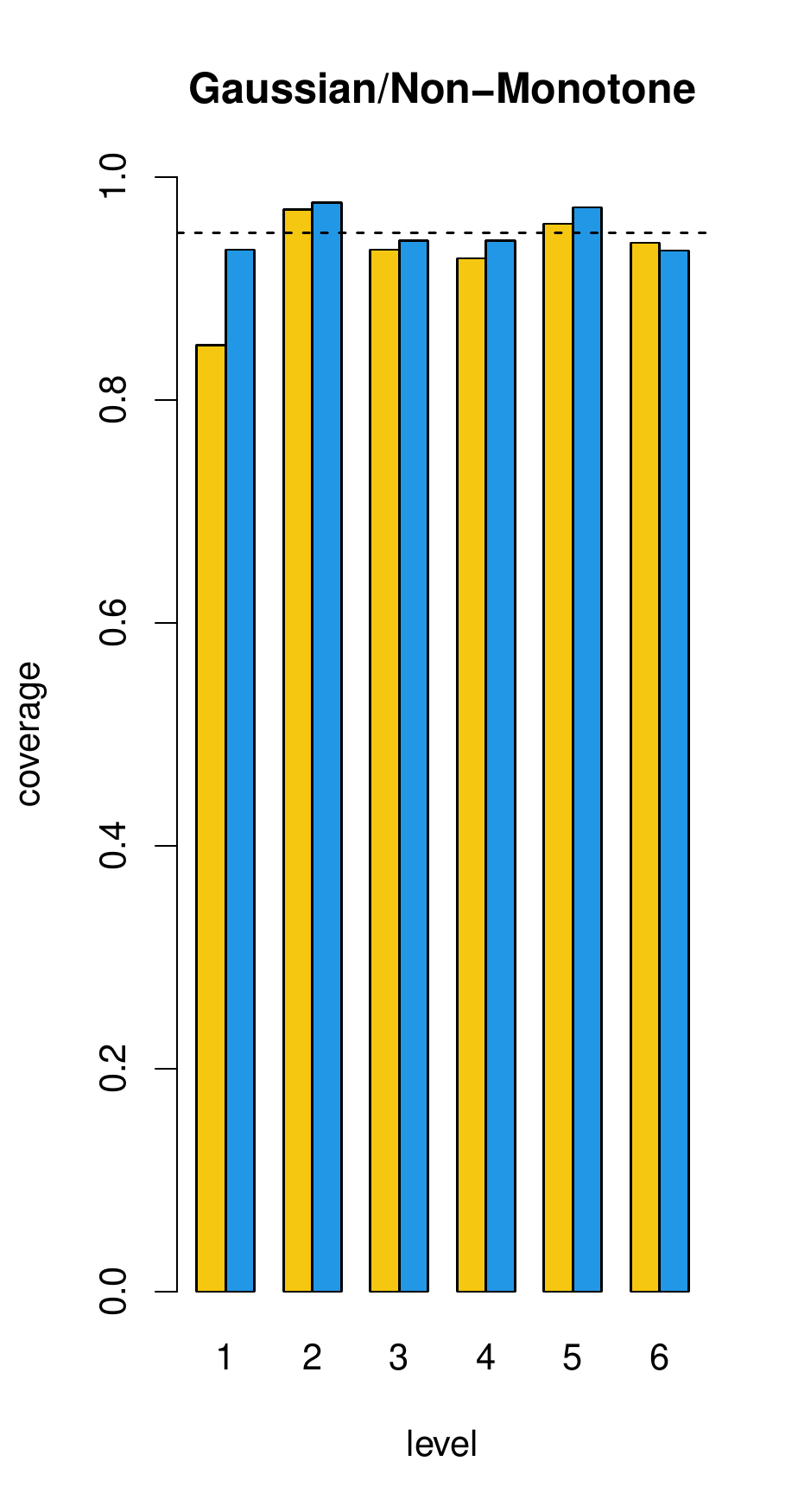}
	\hspace{-4mm}
	\includegraphics[width=45mm]{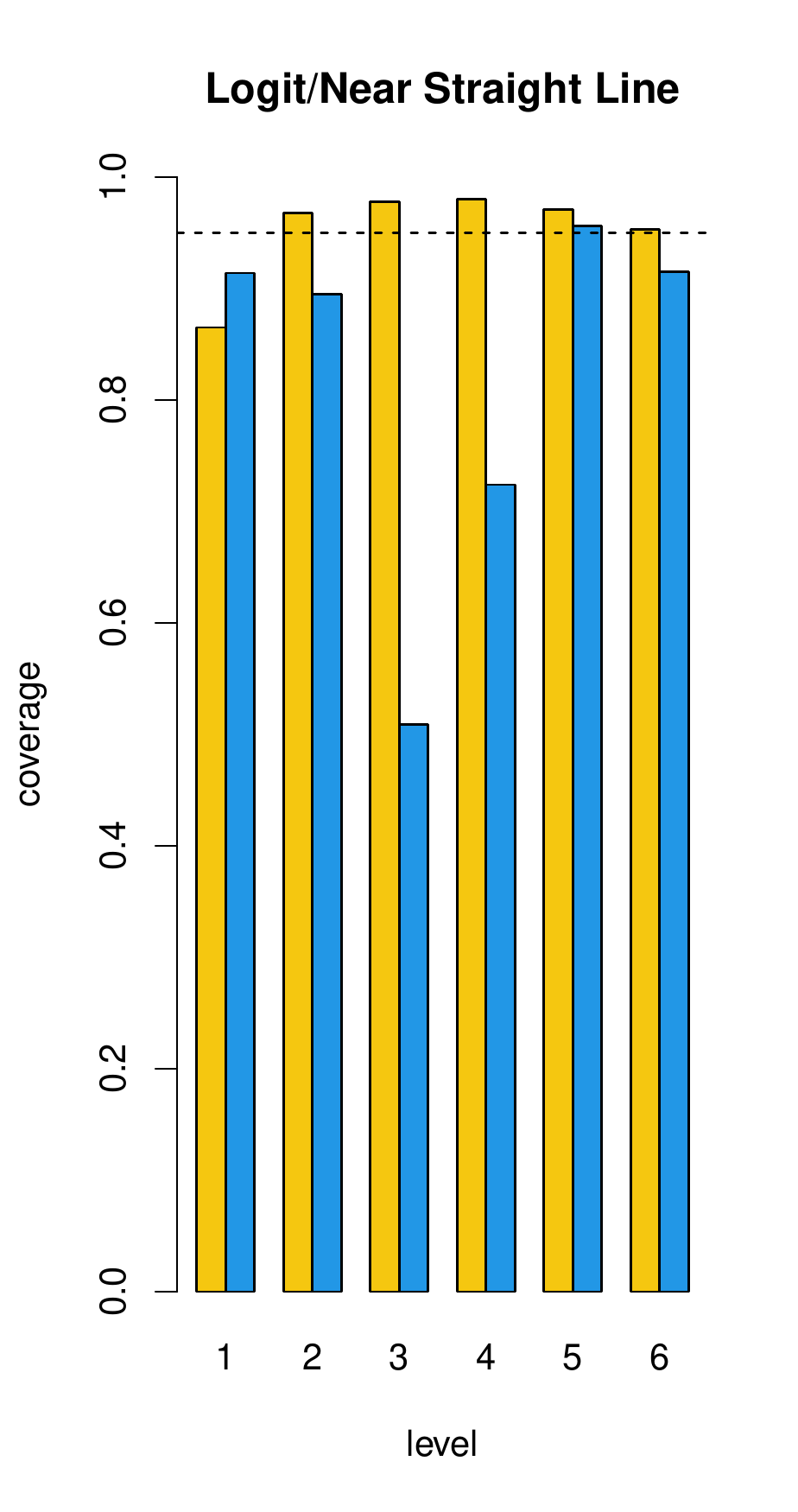}
	\hspace{-4mm}
	\includegraphics[width=45mm]{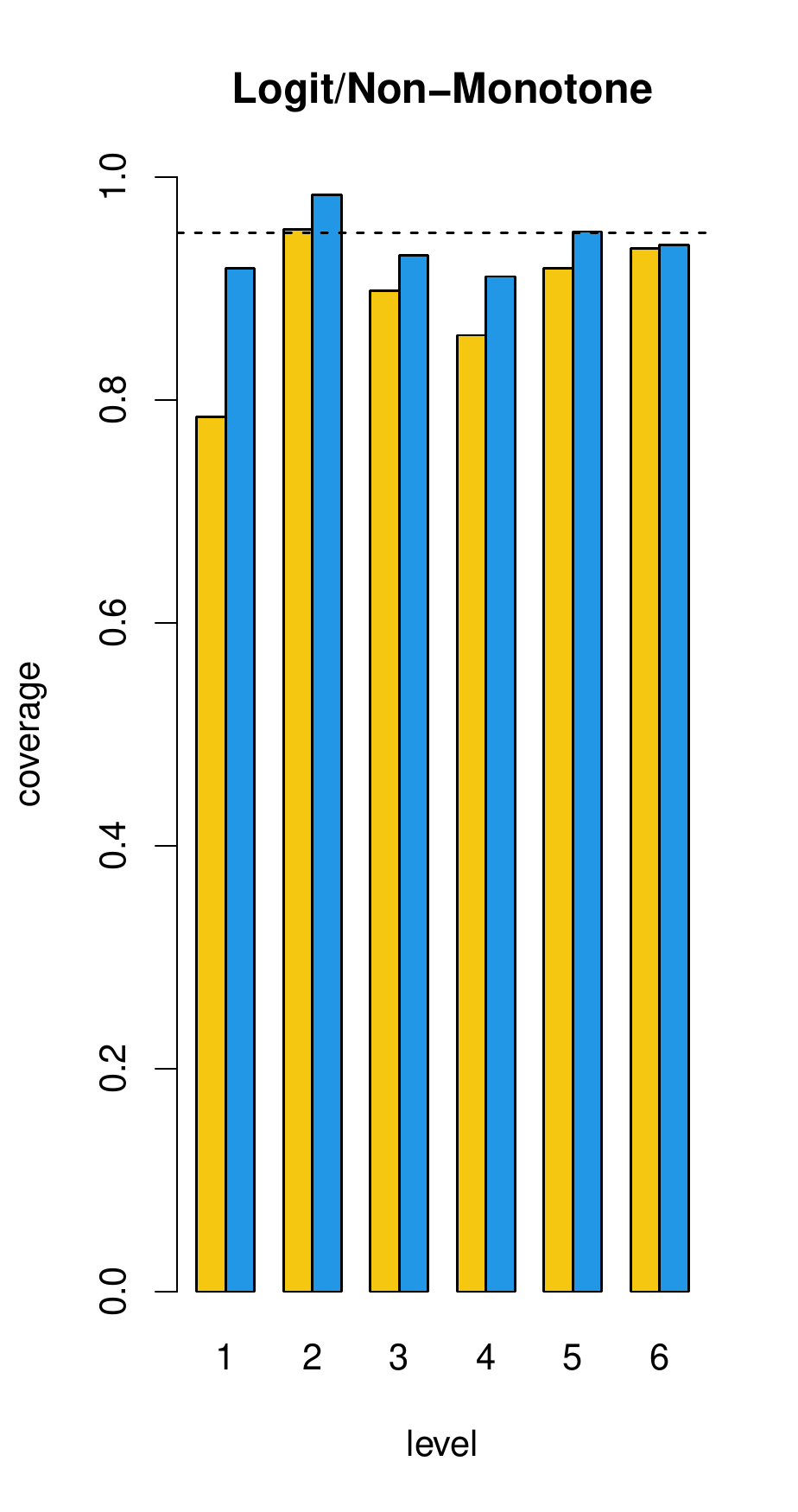}
	\vspace{-3mm}
	\caption{\label{fig:cvrg} Coverage observed for (nominally) 95\% confidence intervals when using the first-order (yellow) and second-order (blue) penalty in different simulation settings.}
\end{figure}

We consider the same settings as in Section~\ref{Sec:SizePower}, with $\psi = 0.8$ and $\psi = 1.6$ in the Gaussian and logit case, respectively. For comparing the estimated to the true functions/coefficients (see Figure~\ref{fig:power}, left), estimated functions/coefficients are centered such that $\sum_l \hat{\beta}_{l} = \sum_l\hat{f}(l) = 0$ (which is also the case for true $f$'s in Figure~\ref{fig:power}). Figure~\ref{fig:cvrg} shows the coverage observed over 1000 replicates for (point-wise) confidence intervals with a nominal level of 95\%. Here we consider both the first- and second-order penalty. Although the tests from Section~\ref{Sec:TestBPD} are not reliable for the first-order penalty, this penalty may still be used for model fitting and confidence intervals provide important information on the estimates' variability.
Indeed, as seen from Figure~\ref{fig:cvrg}, intervals work reasonably well in both the normal and logit case with non-monotone coefficients. If coefficients are close to being linear, coverage (on average) is also close to 95\% when using the first-order penalty. The second-order penalty, however, leads to substantial under-coverage in the `near straight line' case. This problem is also found for (generalized) additive models with continuous covariates if deviations from linearity are penalized and the fitted regression function is close to being linear. The suggested fix is to change the target of inference to the smooth term plus the overall model intercept~\cite{MarWoo2012, Wood2017}.\\

\begin{figure}[tb]\centering
	\includegraphics[width=85mm]{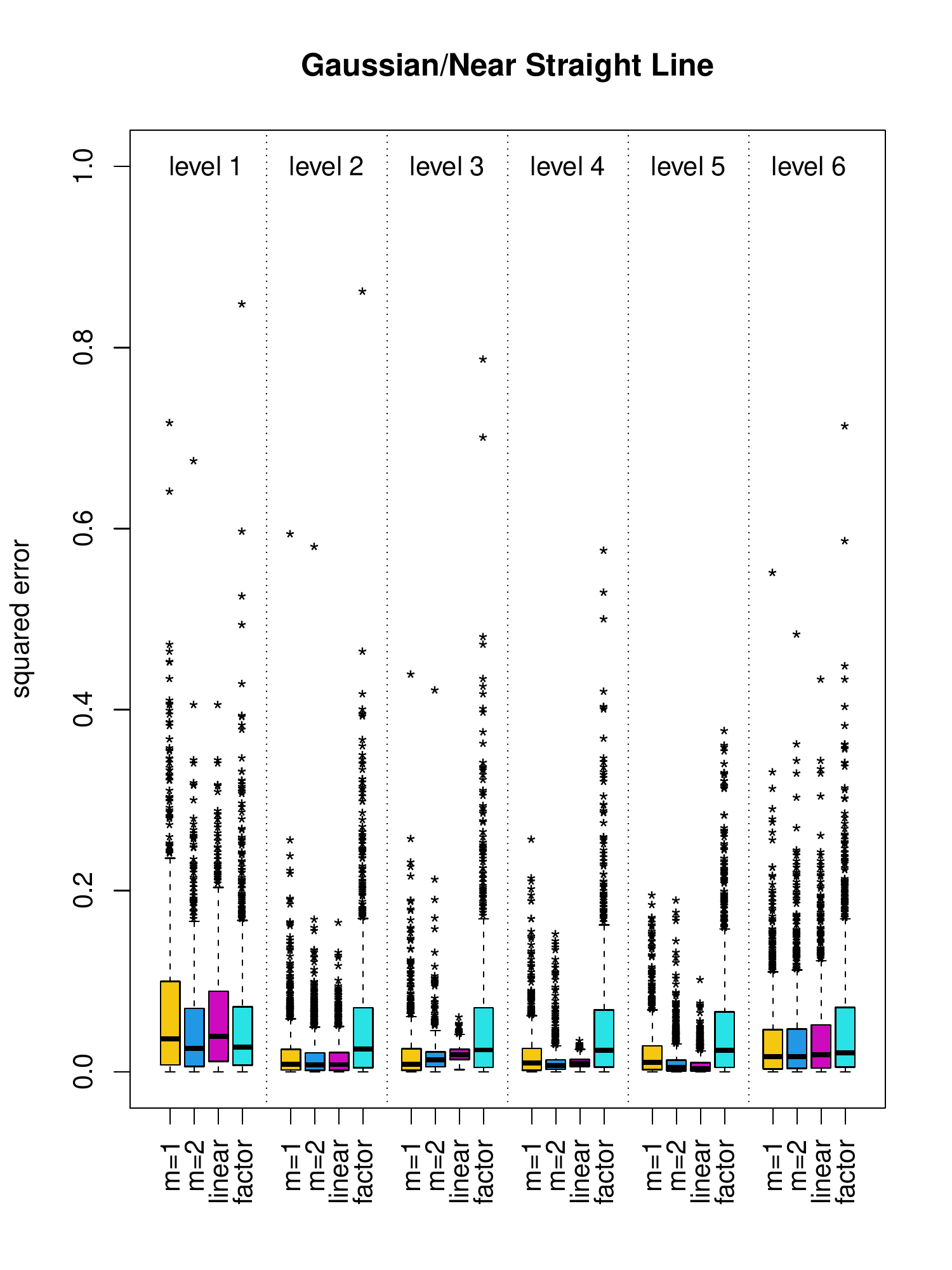}
	\includegraphics[width=85mm]{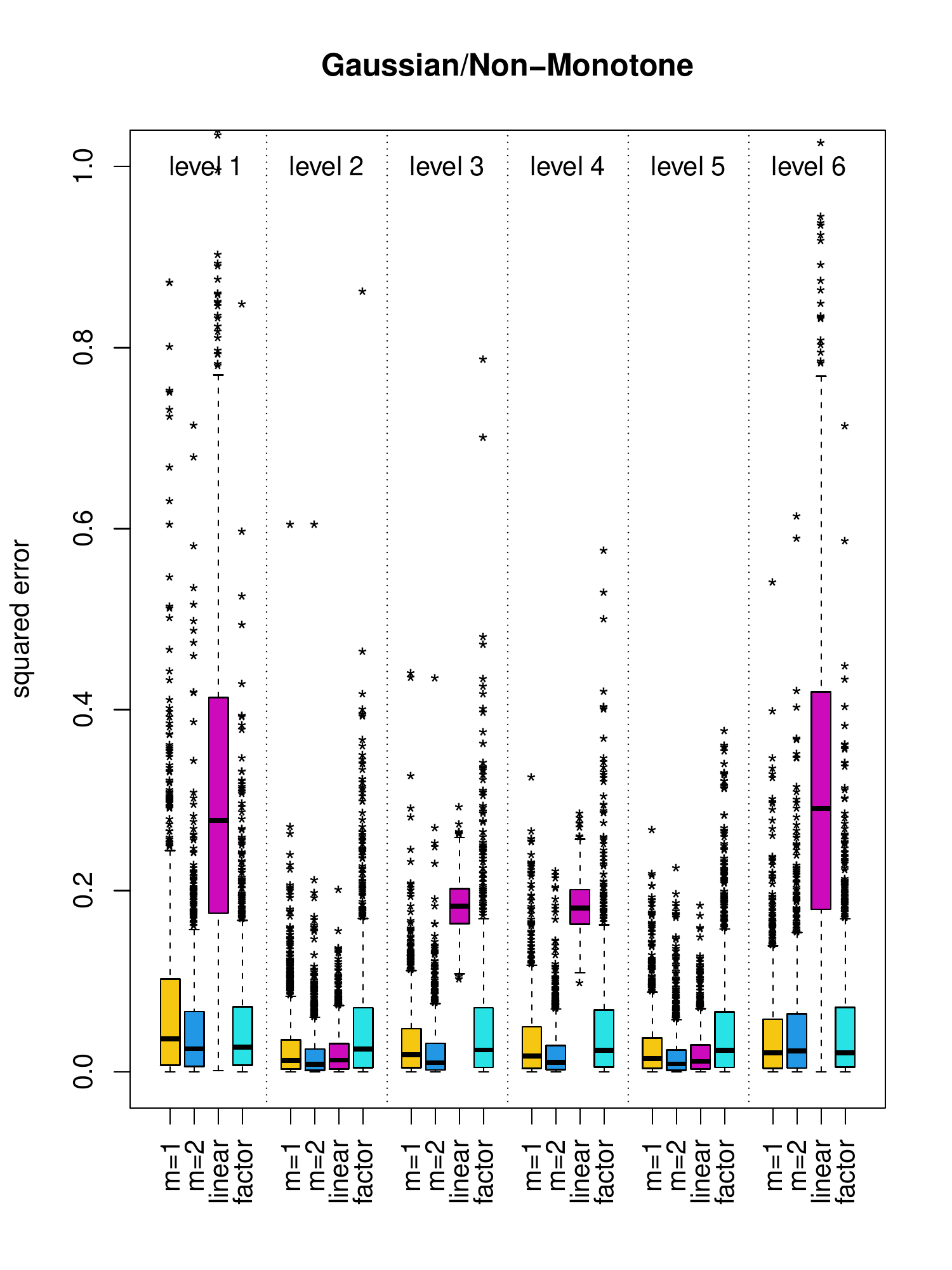}

\vspace{-3mm}
	\caption{\label{fig:mse2} Squared errors in the Gaussian scenarios for first and second-order penalty (i.e., $m=1$ and $m=2$, respectively), linear and ordinary factor modeling; a couple of more extreme outliers are not shown, in particular for ordinary factor modeling.}
\end{figure}

\begin{figure}[tb]\centering
	\includegraphics[width=85mm]{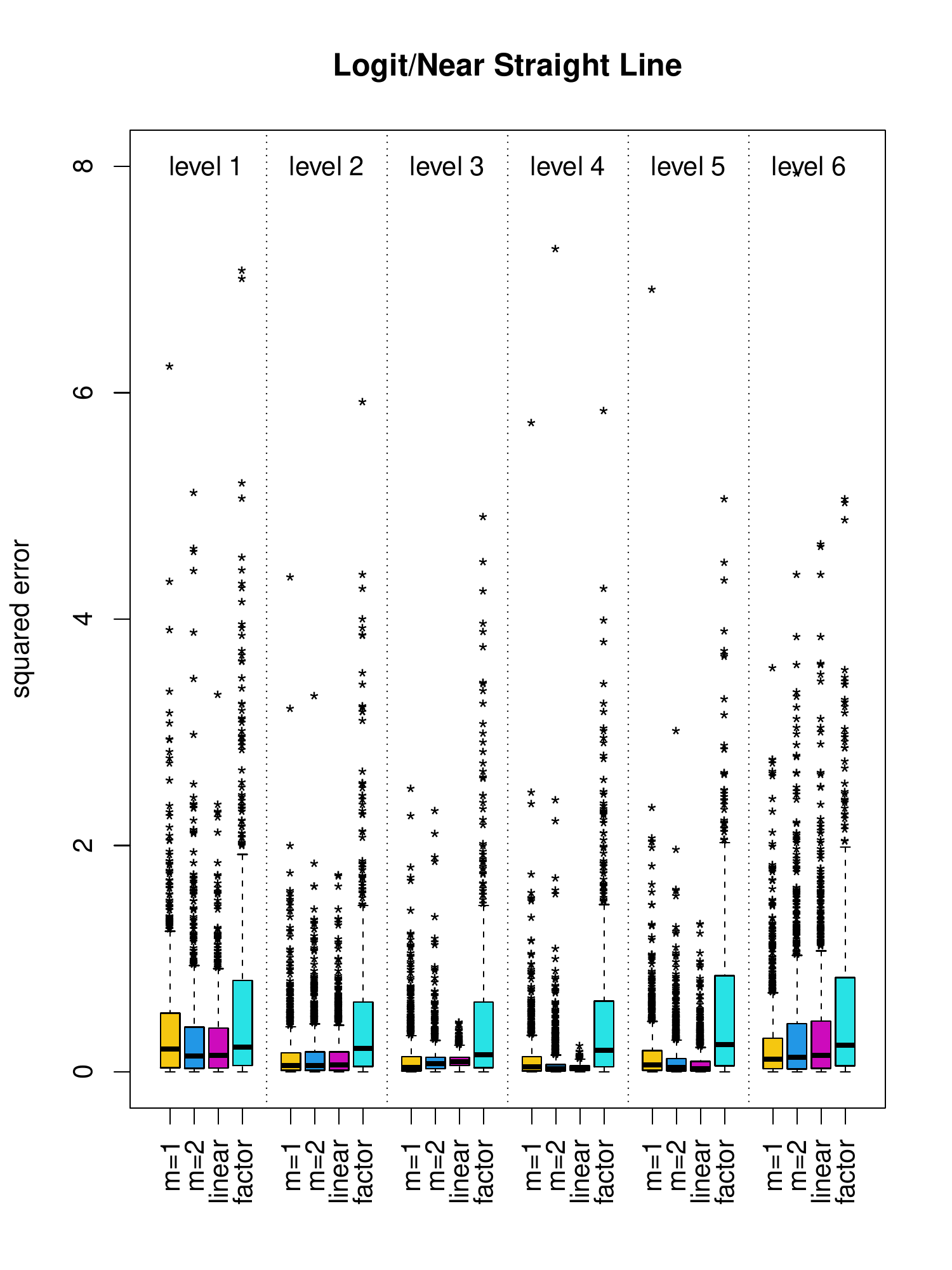}
	\includegraphics[width=85mm]{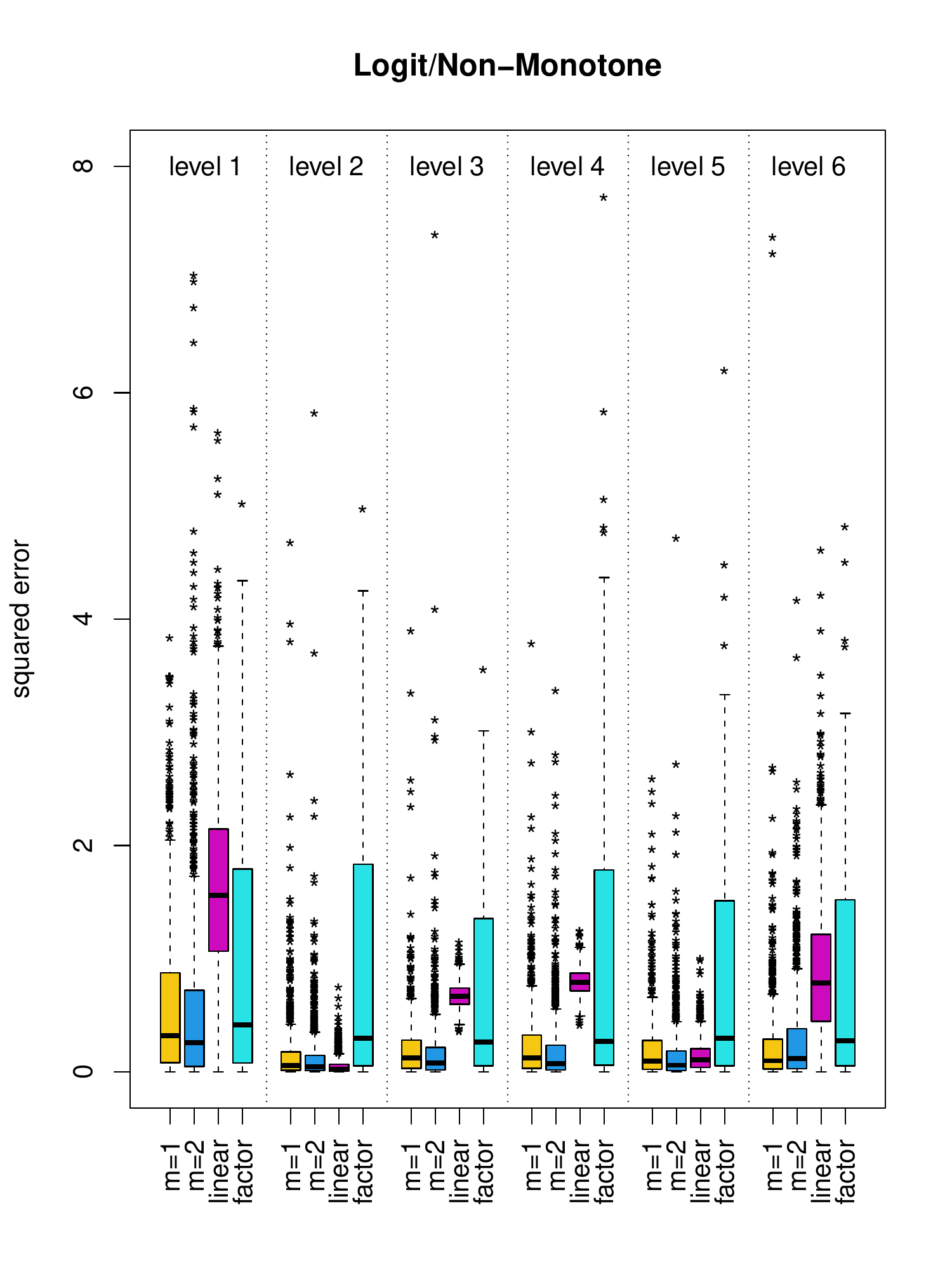}
	
\vspace{-3mm}
	\caption{\label{fig:mse3} Squared errors in the logit scenarios for first and second-order penalty (i.e., $m=1$ and $m=2$, respectively), linear  and ordinary factor modeling; a couple of more extreme outliers are not shown, in particular for ordinary factor modeling.}
\end{figure}

Figure~\ref{fig:mse2} and~\ref{fig:mse3} show graphical summaries of the squared errors when using the first and second-order penalty (i.e., $m=1$ and $m=2$, respectively), linear and ordinary factor modeling. We see that both $m=1$ and $m=2$ produce highly competitive results, with the second-order penalty being the clear overall winner. If the true coefficients nearly lie on a straight line (left panel), assuming linearity over the factor levels also produces good results, which is not surprising of course. In case of the very non-monotone coefficients (right panel), however, the assumption of linearity is highly misleading. Ordinary factor modeling generally suffers from very large variance compared to the penalized estimates, leading to a much larger mean squared error. The is particularly problematic with binary response data (logit model), because sometimes there are only 1s or 0s on some predictor levels, leading to an inflation of the corresponding coefficients, that means, coefficients are tending towards $\pm \infty$. Those very extreme outliers are not even shown in Figure~\ref{fig:mse3}. Of course, those problems may vanish when increasing the sampling size and/or changing the coefficients used to generate the data. With the BPD data, however, we only have a moderate sample size ($n\approx 100$) and even categories with BPD cases/controls only (see Figure~\ref{fig:data}). So we focus on comparable settings here.

\section{Application to Bronchopulmonary Dysplasia}\label{Sec:BPD}

In addition to the model with predictor `overall oral bacterial colonization' as given on page~\pageref{model:overall}, which resulted in a generalized but otherwise purely linear model (due to the `smooth' terms fitted as linear), we also fitted a more detailed model with bacterial colonization categorized into `gram negative', `gram positive', and `pathogenic'. As before, the corresponding ordinal predictor gives the week colonization by the respective type of (oral) bacteria was detected for the first time. In the original publication~\cite{LauEtal2020}, separate models were fit for each type of bacteria, and the first two weeks were collapsed to make (unpenalized) model fitting with dummy-coded ordinal factors feasible. Thanks to the penalties presented here, we are now able to include all three predictors jointly while using all information with the resolution available.

\begin{figure}[tb]\centering

	\includegraphics[width=56mm]{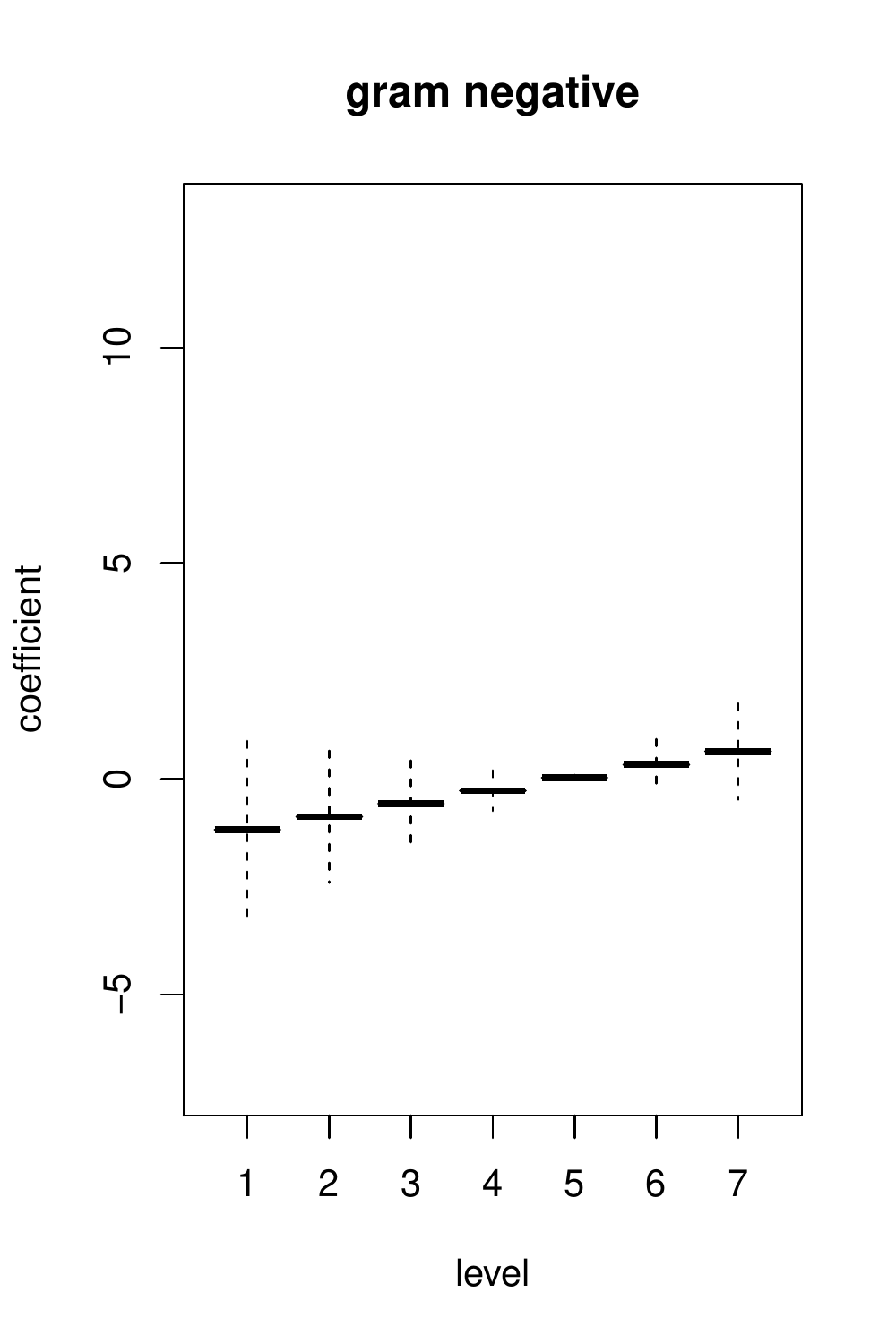}
	\includegraphics[width=56mm]{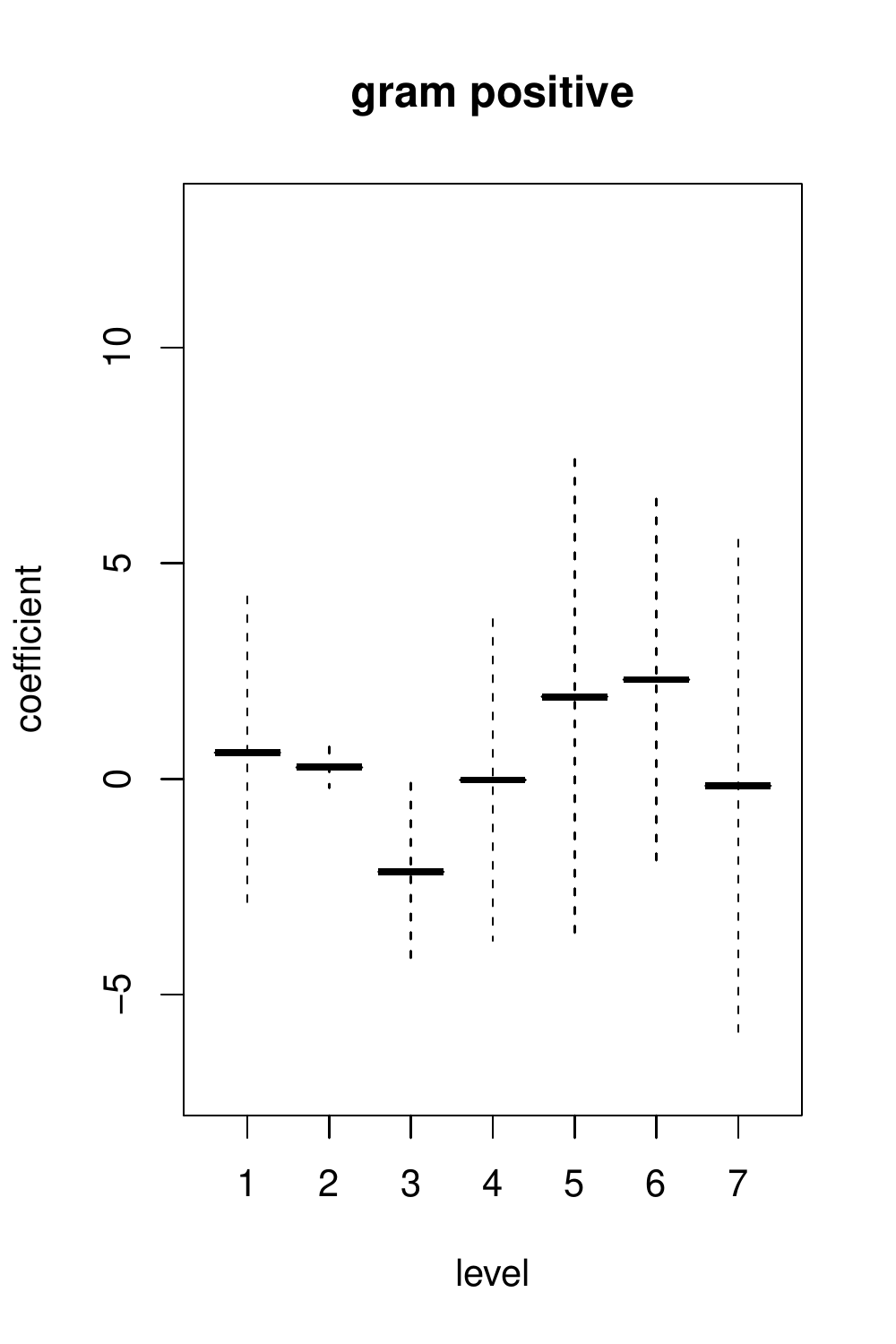}
	\includegraphics[width=56mm]{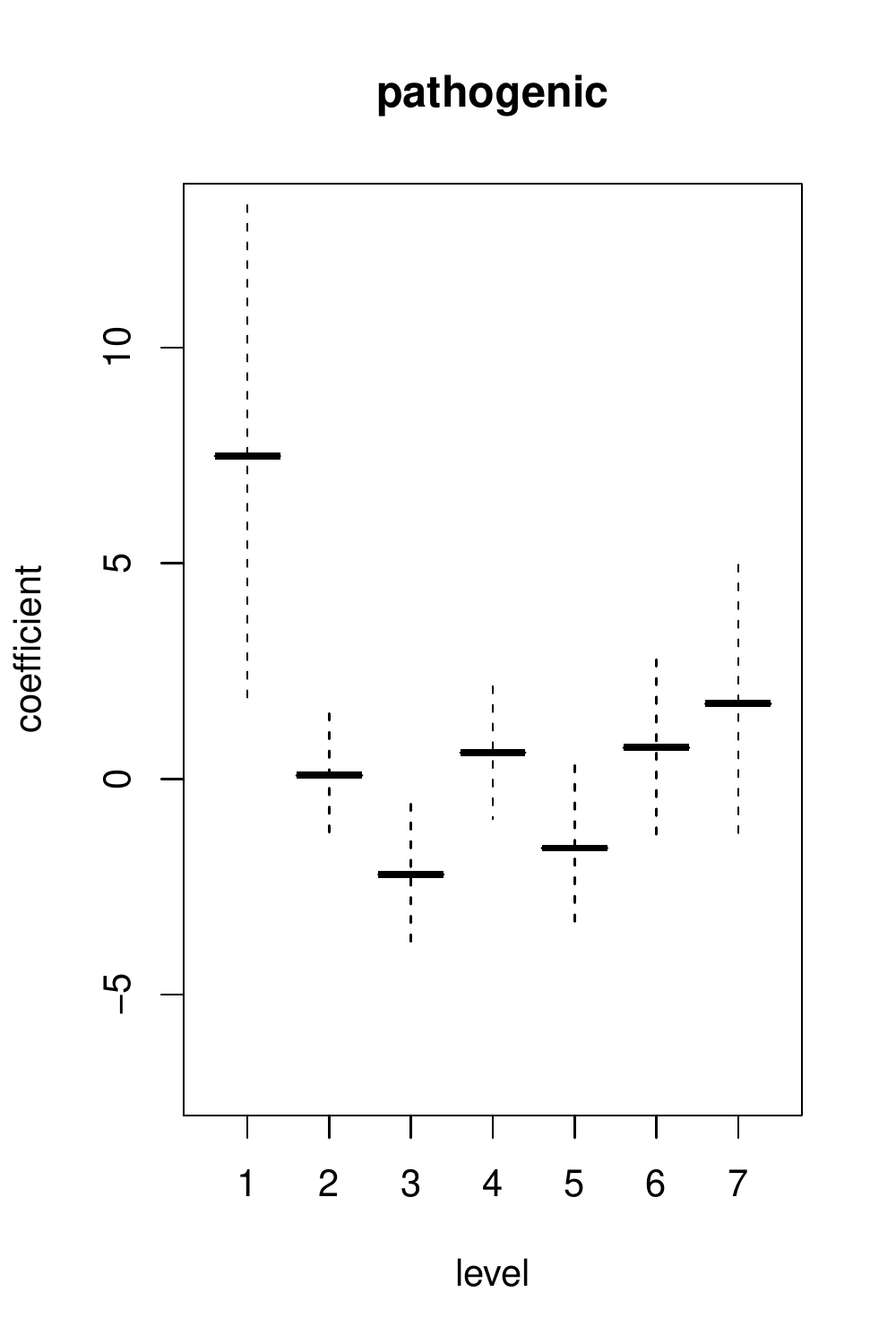}
\vspace{-3mm}
	\caption{\label{fig:coefs} Fitted coefficients of ordinal predictors together with pointwise 95\% confidence intervals in the full model with second-order penalties.}
\end{figure}

Table~\ref{tab:param} (left) shows the results for the parametric terms. In particular it is seen/confirmed that low birth weight is a risk factor for BPD, and also male infants and multiples have an increased risk of developing BPD. Antenatal steroids, by contrast, may decrease the risk. Results for the different types of bacterial colonization, which are included as ordinal predictors with smooth effects, are given in Table~\ref{tab:smooth} and Figure~\ref{fig:coefs}. We see that the only significant effect is detected for pathogenic bacteria. The fitted function indicates that early detection is associated with increased risk of BPD. Statistical uncertainty, however, is very large (due to the small number of samples in category/level 1) as indicated by the confidence interval.
	\begin{table}[t]%
		\begin{center}
		\begin{tabular}{lrrrrrrrrr}
			\toprule
			& \multicolumn{4}{c}{\textbf{Full model}} & & \multicolumn{4}{c}{\textbf{Reduced model}} \\
			\cmidrule{2-5}\cmidrule{7-10}
			\textbf{Covariate} & \textbf{Est.}  & \textbf{S.E}  & \textbf{z-val.}  & \textbf{p-val.} & & \textbf{Est.}  & \textbf{S.E.}  & \textbf{z-val.}  & \textbf{p-val.}  \\
			\midrule
			(Intercept) & 6.214 & 2.630 & 2.363 & 0.018 & & 6.426 & 2.170 & 2.961 & 0.003 \\
			Weight (g) & $-$0.013  & 0.004  & $-$3.381  & $<$0.001 & & $-$0.012 & 0.003 & $-$3.859 & $<$0.001  \\
			SGA sym. &1.909  & 1.359  & 1.405  & 0.160 & & 1.991 & 1.290 & 1.544 & 0.123 \\
			Sex (male) & 3.022 & 1.114 & 2.712 & 0.007 & & 2.107 & 0.834 & 2.527 & 0.012\\
			Multiples & 1.524 & 0.744& 2.048 & 0.041 & & 1.054 & 0.528 & 1.995 & 0.046\\
			Steroids & $-$0.241 & 0.090 & $-$2.684 & 0.007 & & $-$0.174 & 0.072 & $-$2.432 & 0.015\\
			Antibiotics & 0.079 & 0.090 & 0.874 & 0.382 & & 0.079 & 0.078 & 1.013 & 0.311\\
			\bottomrule
		\end{tabular}
			\caption{Results for parametric terms in the full and reduced model when using the second-order ordinal smoothing penalty.\label{tab:param}}
		\end{center}
	\end{table}

When excluding information on gram negative and positive bacteria from the model, results for parametric terms (Table~\ref{tab:param}, right), fitted functions/coefficients for pathogenic bacteria (not shown) and testing results look very similar as before (Table~\ref{tab:smooth}, right). In summary, our results using the ordinal smoothing approach confirm the results obtained earlier~\cite{LauEtal2020}, but allow for considering all three ordinal predictors (gram negative/positive, pathogenic bacteria) jointly. Also the shape of potential effects can be investigated in more detail.

	\begin{table}[b]%
		\begin{center}
\begin{tabular}{lrrrrrrrrr}
			\toprule
			&\multicolumn{4}{c}{\textbf{Full model}} & & \multicolumn{4}{c}{\textbf{Reduced model}} \\
			\cmidrule{2-5}\cmidrule{7-10}
			\textbf{Predictor} & \textbf{edf}  & \textbf{Ref.df}  & \textbf{Chi.sq}  & \textbf{p-val.} & & \textbf{edf}  & \textbf{Ref.df}  & \textbf{Chi.sq}  & \textbf{p-val.}  \\
			\midrule
			Gram negative & 1.000 & 1.000 & 1.307 & 0.253 & & -- & -- & -- & -- \\
			Gram positive & 3.708  & 4.393  & 5.227  & 0.264 & & -- & -- & -- & --  \\
			Pathogenic & 5.258  & 5.831  & 13.711  & 0.030 & & 4.973 & 5.696 & 13.573 & 0.027 \\
			\bottomrule
		\end{tabular}
			\caption{Results for smooth terms in the full and reduced model when using the second-order ordinal smoothing penalty.\label{tab:smooth}}
	\end{center}
	\end{table}

\section{Concluding Remarks}\label{Sec:Conclude}

We showed how ordinal predictors in generalized linear models that are fitted by use of quadratic difference penalties as proposed in the literature can be incorporated and interpreted in the framework of generalized additive models, making tools for statistical inference developed there available for ordinal predictors as well. Here we focused on fitting itself with built-in selection of penalty parameters, and further statistical inference in terms of (pointwise) confidence intervals and testing significance of smooth terms, i.e., significance of the ordinal predictors. In particular, we confirmed that at least some of the tests provided in the popular \texttt{mgcv} R package can be used with ordinal predictors, too. More specifically, this is the case for the Wald-type tests when employing the second-order difference penalty on adjacent dummy coefficients. It should be pointed out, however, that this is particularly true if smoothing parameters are estimated by REML (or ML), which has been investigated in this study. When using GCV (which is the default in \texttt{mgcv}), this is not necessarily the case. 

Besides testing in a pre-defined model as done here, those tests may also be used for stepwise (backward/forward) variable selection, which is still quite popular among applied researchers, including but not limited to medicine. However, the methods discussed here are primarily useful if the number of (ordinal) covariates is moderate. In high-dimensional problems, more precisely, if the number of covariates becomes large, other techniques than statistical testing may be used for model selection; for instance, sparsity inducing penalties like the group lasso~\cite{YuaLin2006, GerEtal2011}. 

Add-on functions implementing the ordinal basis for use within \texttt{mgcv} are publicly available through R package \texttt{ordPens}~\cite{GerHos2021}. After installing and loading \texttt{ordPens}, the \texttt{gam()} function from \texttt{mgcv} can be used with smooth terms \texttt{s(..., bs = "ordinal", m = 1)} or \texttt{s(..., bs = "ordinal", m = 2)} for the first- and second-order penalty, respectively.

In summary, using the ordinal smoothing penalty within generalized additive models offers a very convenient and flexible way to respect and exploit the information provided by ordinally scaled predictors in a sound statistical framework with elaborated tools for statistical inference. Built-in selection of tuning parameters yields estimates that adapt very well to the data structure with accuracy at least competitive if not better than standard linear/factor modeling.

\subsection*{Author contributions}

Jan Gertheiss conducted the data analysis/numerical experiments and wrote the manuscript. Fabian Scheipl implemented the method for use within \texttt{mgcv}. Tina Lauer and Harald Ehrhardt collected the data and were involved in the discussion of the manuscript.

\subsection*{Funding}

This work was supported in part by Deutsche Forschungsgemeinschaft (DFG) under Grant GE2353/2-1.

\bibliographystyle{plain}
\bibliography{arxivGAMordinal}%

\begin{thebibliography}{10}

\bibitem{BreCla1993}
N~E Breslow and D~G Clayton.
\newblock Approximate inference in generalized linear mixed models.
\newblock {\em Journal of the American Statistical Association}, 88:9--25,
  1993.

\bibitem{CieEtal2014}
A~Cieza, C~Oberhauser, J~Bickenbach, S~Chatterji, and G~Stucki.
\newblock Towards a minimal generic set of domains of functioning and health.
\newblock {\em BMC Public Health}, 14:218, 2014.

\bibitem{CraRup2004}
C~M Crainiceanu and D~Ruppert.
\newblock Likelihood ratio tests in linear mixed models with one variance
  component.
\newblock {\em Journal of the Royal Statistical Society B}, 66:165--185, 2004.

\bibitem{CraEtal2005}
C~M Crainiceanu, D~Ruppert, G~Claeskens, and M~P Wand.
\newblock Exact likelihood ratio tests for penalized splines.
\newblock {\em Biometrika}, 77:91--103, 2005.

\bibitem{deBoor1978}
C~{de Boor}.
\newblock {\em A Practical Guide to Splines}.
\newblock Springer, New York, 1978.

\bibitem{Dierckx1993}
P~Dierckx.
\newblock {\em Curve and Surface Fitting with Splines}.
\newblock Claredon Press, Oxford, 1993.

\bibitem{EilMar1996}
P~H~C Eilers and B~D Marx.
\newblock Flexible smoothing with {B}-splines and penalties.
\newblock {\em Statistical Science}, 11:89--121, 1996.

\bibitem{EilMar2002}
P~H~C Eilers and B~D Marx.
\newblock Generalized linear additive smooth structures.
\newblock {\em Journal of Computational and Graphical Statistics}, 11:758--783,
  2002.

\bibitem{FahEtal2013}
L~Fahrmeir, T~Kneib, S~Lang, and B~Marx.
\newblock {\em Regression--Models, Methods and Applications}.
\newblock Springer, Berlin/Heidelberg, 2013.

\bibitem{Ger2014}
J~Gertheiss.
\newblock Anova for factors with ordered levels.
\newblock {\em Journal of Agricultural, Biological, and Environmental
  Statistics}, 19:258--277, 2014.

\bibitem{GerEtal2011}
J~Gertheiss, S~Hogger, C~Oberhauser, and G~Tutz.
\newblock Selection of ordinally scaled independent variables with applications
  to international classification of functioning core sets.
\newblock {\em Journal of the Royal Statistical Society C}, 60:377--395, 2011.

\bibitem{GerHos2021}
J~Gertheiss and A~Hoshiyar.
\newblock {\em ordPens: Selection, Fusion, Smoothing and Principal Components
  Analysis for Ordinal Variables}, 2021.
\newblock R package version 1.0.0.

\bibitem{GerOeh2011}
J~Gertheiss and F~Oehrlein.
\newblock Testing relevance and linearity of ordinal predictors.
\newblock {\em Electronic Journal of Statistics}, 5:1935--1959, 2011.

\bibitem{GerTut2009}
J~Gertheiss and G~Tutz.
\newblock Penalized regression with ordinal predictors.
\newblock {\em International Statistical Review}, 77:345--365, 2009.

\bibitem{GlaRos2015}
S~M Glass and S~E Ross.
\newblock Modified functional movement screening as a predictor of tactical
  performance potential in recreationally active adults.
\newblock {\em The International Journal of Sports Physical Therapy},
  10:612--621, 2015.

\bibitem{GreEtal2008}
S~Greven, C~M Crainiceanu, H~K\"uchenhoff, and A~Peters.
\newblock Restricted likelihood ratio testing for zero variance components in
  linear mixed models.
\newblock {\em Journal of Computational and Graphical Statistics}, 17:870--891,
  2008.

\bibitem{GroEtal2018}
J~Gronbach, T~Shahzad, S~Radajewski, C-M Chao, S~Bellusci, R~E Morty,
  T~Reicherzer, and H~Ehrhardt.
\newblock The potentials and caveats of mesenchymal stromal cell-based
  therapies in the preterm infant.
\newblock {\em Stem Cells International}, page Article ID 9652897, 2018.

\bibitem{Har1974}
D~A Harville.
\newblock Bayesian inference for variance components using only error
  contrasts.
\newblock {\em Biometrika}, 61:383--385, 1974.

\bibitem{Har1977}
D~A Harville.
\newblock Maximum likelihood approaches to variance component estimation and to
  related problems.
\newblock {\em Journal of the American Statistical Association}, 72:320--338,
  1977.

\bibitem{HasTib1990}
T~Hastie and R~Tibshirani.
\newblock {\em Generalized Additive Models}.
\newblock Chapman \& Hall, London, 1990.

\bibitem{HoeKen1970}
A~E Hoerl and R~W Kennard.
\newblock Ridge regression: Biased estimation for nonorthogonal problems.
\newblock {\em Technometrics}, 12:55--67, 1970.

\bibitem{LaiWar1982}
N~M Laird and J~H Ware.
\newblock Random-effects models for longitudinal data.
\newblock {\em Biometrics}, 38:963--974, 1982.

\bibitem{LauEtal2020}
T~Lauer, J~Behnke, F~Oehmke, J~Bäcker, K~Gentil, T~Chakraborty, M~Schloter,
  J~Gertheiss, and H~Ehrhardt.
\newblock Bacterial colonization within the first six weeks of life and
  pulmonary outcome in preterm infants < 1000g.
\newblock {\em Journal of Clinical Medicine}, 9:2240, 2020.

\bibitem{MarWoo2012}
G~Marra and S~N Wood.
\newblock Coverage properties of confidence intervals for generalized additive
  model components.
\newblock {\em Scandinavian Journal of Statistics}, 39:53--74, 2012.

\bibitem{MarEil1998}
B~D Marx and P~H~C Eilers.
\newblock Direct generalized additive modelling with penalized likelihood.
\newblock {\em Computational Statistics \& Data Analysis}, 28:193--209, 1998.

\bibitem{McCNel1989}
P~McCullagh and J~A Nelder.
\newblock {\em Generalized Linear Models}.
\newblock Chapman \& Hall, New York, 2nd edition, 1989.

\bibitem{NelWed1972}
J~A Nelder and R~W~M Wedderburn.
\newblock Generalized linear models.
\newblock {\em Journal of the Royal Statistical Society A}, 135:370--384, 1972.

\bibitem{Nychka1988}
D~Nychka.
\newblock Bayesian confidence intervals of smoothing splines.
\newblock {\em Journal of the American Statistical Association}, 83:1134--1143,
  1988.

\bibitem{PatTho1971}
H~D Patterson and R~Thompson.
\newblock Recovery of interblock information when block sizes are unequal.
\newblock {\em Biometrika}, 58:545--554, 1971.

\bibitem{RCore2020}
{R Core Team}.
\newblock {\em R: A Language and Environment for Statistical Computing}.
\newblock R Foundation for Statistical Computing, Vienna, Austria, 2020.

\bibitem{SchGreKue2008}
F~Scheipl, S~Greven, and H~K\"uchenhoff.
\newblock Size and power of tests for a zero random effect variance or
  polynomial regression in additive and linear mixed models.
\newblock {\em Computational Statistics \& Data Analysis}, 52:3283--3299, 2008.

\bibitem{SweCraGer2016}
E~Sweeney, C~Crainiceanu, and J~Gertheiss.
\newblock Testing differentially expressed genes in dose-response studies and
  with ordinal phenotypes.
\newblock {\em Statistical Applications in Genetics and Molecular Biology},
  15:213--235, 2016.

\bibitem{TutGer2014}
G~Tutz and J~Gertheiss.
\newblock Rating scales as predictors -- the old question of scale level and
  some answers.
\newblock {\em Psychometrika}, 79:357--736, 2014.

\bibitem{TutGer2016}
G~Tutz and J~Gertheiss.
\newblock Regularized regression for categorical data (with discussion).
\newblock {\em Statistical Modelling}, 16:161--200, 2016.

\bibitem{Wood2008}
S~N Wood.
\newblock Fast stable direct fitting and smoothness selection for generalized
  additive models.
\newblock {\em Journal of the Royal Statistical Society B}, 70:495--518, 2008.

\bibitem{Wood2011}
S~N Wood.
\newblock Fast stable restricted maximum likelihood and marginal likelihood
  estimation of semiparametric generalized linear models.
\newblock {\em Journal of the Royal Statistical Society B}, 73:3--36, 2011.

\bibitem{Wood2013}
S~N Wood.
\newblock On p-values for smooth components of an extended generalized additive
  model.
\newblock {\em Biometrika}, 100:221--228, 2013.

\bibitem{Wood2017}
S~N Wood.
\newblock {\em Generalized Additive Models: An Introduction with R}.
\newblock CRC Press, Boca Raton, 2nd edition, 2017.

\bibitem{WooPyaSae2016}
S~N Wood, N~Pya, and B.~Saefken.
\newblock Smoothing parameter and model selection for general smooth models
  (with discussion).
\newblock {\em Journal of the American Statistical Association},
  111:1548--1575, 2016.

\bibitem{YuaLin2006}
M~Yuan and Y~Lin.
\newblock Model selection and estimation in regression with grouped variables.
\newblock {\em Journal of the Royal Statistical Society B}, 68:49--67, 2006.

\end{thebibliography}

\end{document}